\documentclass[12pt,a4paper]{article}
\usepackage{amssymb,enumitem,booktabs,subfig,xcolor,todonotes,microtype,setspace,bm,paralist, dsfont, longtable} 
\usepackage[top=1in, bottom=1in, left=1in, right=1in]{geometry}
\usepackage[numbers]{natbib}
\usepackage{url}
\usepackage[pdftex,colorlinks=true,hypertexnames=false]{hyperref}
\definecolor{darkblue}{rgb}{0,0,.6}
\hypersetup{citecolor=darkblue,linkcolor=darkblue,urlcolor=darkblue}
\usepackage{amsmath}
\usepackage{amsfonts}
\usepackage{epsfig}
\usepackage{graphics}
\usepackage{palatino,eulervm}
\usepackage{graphicx}
\usepackage{lscape}
\usepackage{rotating}
\usepackage[linewidth=1pt]{mdframed}
\usepackage{orcidlink}
\allowdisplaybreaks[4]
\DeclareMathOperator*{\argmin}{arg\,min}

\newcommand\X{\mathcal{X}}

\providecommand{\U}[1]{\protect\rule{.1in}{.1in}}

\graphicspath{{plots/}}

\usepackage{amsthm,thmtools}
\usepackage{mathrsfs}

\makeatletter
\def\th@newremark{\th@remark\thm@headfont{\bfseries}}
\makeatletter
\theoremstyle{newremark}

\declaretheoremstyle[
  spaceabove=6pt, spacebelow=6pt,
  headfont=\bfseries,
  notefont=\mdseries, notebraces={(}{)},
bodyfont=\normalfont,
  postheadspace=0.5em]{mystyle}

\newsavebox\CBox
\def\textBF#1{\sbox\CBox{#1}\resizebox{\wd\CBox}{\ht\CBox}{\textbf{#1}}}
\newcommand{\Rlogo}{\protect\includegraphics[height=1.8ex,keepaspectratio]{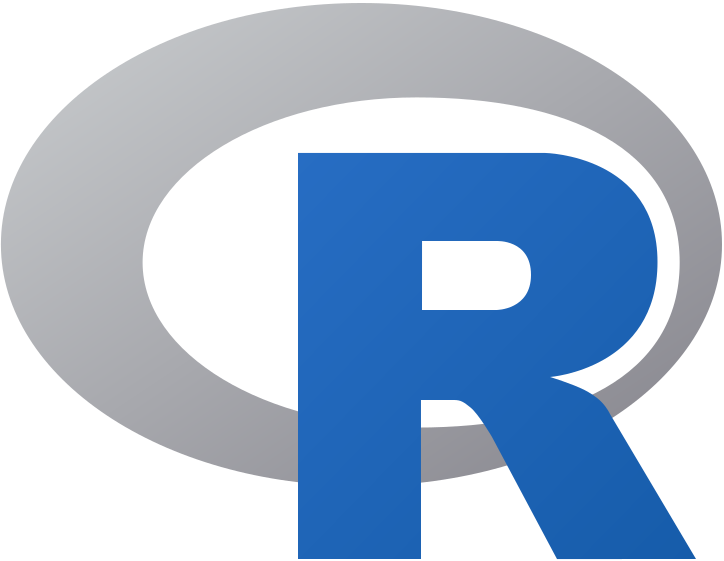}}

\begin{document}

\title{Forecasting a time series of Lorenz curves: \mbox{One-way functional analysis of variance}}
\author{{\normalsize Han Lin Shang\orcidlink{0000-0003-1769-6430}\thanks{Postal address: Department of Actuarial Studies and Business Analytics, Macquarie University, Sydney, NSW 2109, Australia. Telephone: +61(2) 9850 4689; Email: hanlin.shang@mq.edu.au}}\\
\normalsize Macquarie University}

\date{}

\maketitle

\centerline{\bf Abstract}

\medskip

The Lorenz curve is a fundamental tool for analysing income and wealth distribution and inequality at national and regional levels. We utilise a one-way functional analysis of variance to decompose a time series of Lorenz curves and develop a method for producing one-step-ahead point and interval forecasts. The one-way functional analysis of variance is easily interpretable by decomposing an array into a functional grand effect, a functional row effect and residual functions. We evaluate and compare the forecast accuracy between the functional analysis of variance and three non-functional methods using the Italian household income and wealth data.

\medskip


\setstretch{1.45}
\newpage

\section{Introduction}\label{sec:1}

The Lorenz curve \citep{Lorenz05} is a fundamental tool for analysing income and wealth distribution and identifying regional inequality. Mathematically, the Lorenz curve is non-decreasing and convex, with $L(0) = 0$ and $L(1)=1$. Given the constraints, the Lorenz curve shares a strong resemblance to a cumulative distribution function (CDF). The Lorenz curve can be defined as:
\begin{equation*}
L(p) := \frac{\int_0^pQ(\tau)d\tau}{\int^1_0Q(\tau)d\tau}, \quad p\in[0,1],
\end{equation*}
where $F(q)$ is the CDF and $Q(p) = \inf\{q\in R|F(q)\geq p\}$ is the quantile function. It can be interpreted as the cumulative share of the income accumulated by the bottom $p$ proportion of the population. For example, if the poorest $80\%$ of the households in a society hold 20\% of the income, we have $L(0.8)=0.2$. Being a continuous function, the Lorenz curve and its derivative provide important information regarding inequality and give insights into how income is distributed in a society. 

Based on the Lorenz curve, its derivative is called share density \citep{Farris10, Zizler14}, 
\begin{equation*}
\frac{dL}{dp}=\vartheta(p), 
\end{equation*}
to indicate the relation with the share of total income owned by a small portion of a population, where $p$ refers to the fraction of the population that holds $L(p)$ proportion of the whole income. In a society with equally distributed income, we observe the constant share density function $\vartheta(p)=1$ for all $p$. 

The expected value of the share density introduces a concept of the percentile level of a household, which earns the average dollar. It can be expressed as
\begin{equation*}
\overline{p}=\int^{1}_{0} p\vartheta(p)dp.
\end{equation*}

The Lorenz curve is also a key part of the calculation for the Gini index \citep{Gini36}, which is a single number measuring how the income is spread in a population equitably. The Gini index can be expressed as
\begin{equation*}
G = 2\int^{1}_{0}\left[p-L(p)\right]dp.
\end{equation*}
The Gini index can also be considered as a measure of health inequality \citep[see, e.g.,][]{LHR+08} or mortality inequality \citep[see, e.g.,][]{SHX22}. The connection between $\overline{p}$ and G is that $G=2\overline{p}-1$. 

The Lorenz curve is a natural example of functional data, displayed in a graphical form of curves, images or shapes. Functional data analysis is collected in monographs of \cite{RS05}, \cite{FV06} and \cite{HK12}. The ability to consider derivatives, a by-product of conceiving the data as functions, is an advantage for visualisation \citep{Shang19} and modelling \citep{HS22}. It also gives rise to dynamic data analysis in \cite{RH17} and \cite{HLW+24}. 

From a policy decision-making perspective, it is crucial to understand income disparities across countries and regions, socioeconomic status, and ethnic groups \citep[see, e.g.,][]{HY10}, and understand its underlying dynamics from historical observations. Because of the availability of subnational data in Italy, \cite{Condino23} used the most recent survey to determine groups of earners by measuring the similarity between the Lorenz curves and their derivatives using a proper similarity measure. 

We are interested in the time series of the Lorenz curves constructed for various regions in Italy, which can be viewed as high-dimensional functional time series (HDFTS). In the HDFTS literature,  \cite{TSY22} considered the problem of clustering multiple functional time series, while \cite{ZD23} studied statistical inference for functional panel data. \cite{GSY19} presented a modelling and forecasting method for HDFTS, while \cite{LLS24} studied the change point detection for identifying common change points. For modelling HDFTS, \cite{TNH23} presented a functional factor model with functional loadings and scalar factors and \cite{GQW22} presented another functional factor model with scalar loading and functional factors.

We contribute to the modelling and forecasting of HDFTS by considering a one-way functional analysis of variance (ANOVA) (see, e.g., \cite{Zhang13} for review). Via the functional ANOVA, we extract the functional grand effect, functional row effect (measuring the variation among regions), and residual functions. While the functional grand and row effects are deterministic, the residual functions are time-varying. For modelling and forecasting the residual functions, we consider a functional time series forecasting method based on a functional principal component analysis (FPCA). A set of residual functions can be approximated by functional principal components and their associated uncorrelated scores. By extrapolating the scores, we obtain $h$-step-ahead forecast residual functions conditional on the estimated mean and estimated functional principal components. The forecast curves can be obtained by adding the forecast residual functions with the functional grand and row effects. 

The outline of this paper is described as follows: In Section~\ref{sec:2}, we describe a motivating data set, i.e., the Italian household income and wealth dataset. From the household income data, we can construct the Lorenz curve via linear interpolation and study its patterns over 11 years from 1998 to 2020. Our functional data are densely and regularly observed, a linear interpolation algorithm of \cite{Hyndman24} can adequately recover whole curves \citep[see, e.g.,][]{ZW16}.

Since the Lorenz curves resemble the CDF, we transform the data via the logit transformation, where the transformed data lie in a real line. To model these transformed data, we revisit a univariate functional time series forecasting method in Sections~\ref{sec:3.1} and~\ref{sec:3.2}. The method captures the temporal dependence in a region but ignores the potential spatial effect. To rectify this problem, we introduce one-way functional ANOVA in Section~\ref{sec:3.3}. In Section~\ref{sec:4}, we present out-of-sample forecasting results and evaluate and compare the forecast accuracy with some holdout data. Section~\ref{sec:5} concludes with some ideas on how the methodology proposed can be further extended.

\section{Italian household income and wealth dataset}\label{sec:2}

We aim to understand how the Lorenz curves vary by region in Italy. Sourced from the Bank of Italy (\footnotesize \url{https://www.bancaditalia.it/statistiche/tematiche/indagini-famiglie-imprese/index.html})\normalsize, we consider a secondary data set from the survey on household income and wealth. This survey includes income, wealth, and other aspects of the economic and financial situation of around 8,000 Italian households from 20 regions. The one-way functional ANOVA can model spatial dependence. Differing from \cite{Condino23}, we study how the Lorenz curves in each region vary over time from 1998 to 2020. The Lorenz curves are observed from 1998 to 2020 every two years, except for the missing year, 2018. There are 11 curves in total for each region.
\begin{figure}[!htb]
\centering
\subfloat[Averaged logit transformation of Lorenz curves over regions (time dimension)]
{\includegraphics[width=8.5cm]{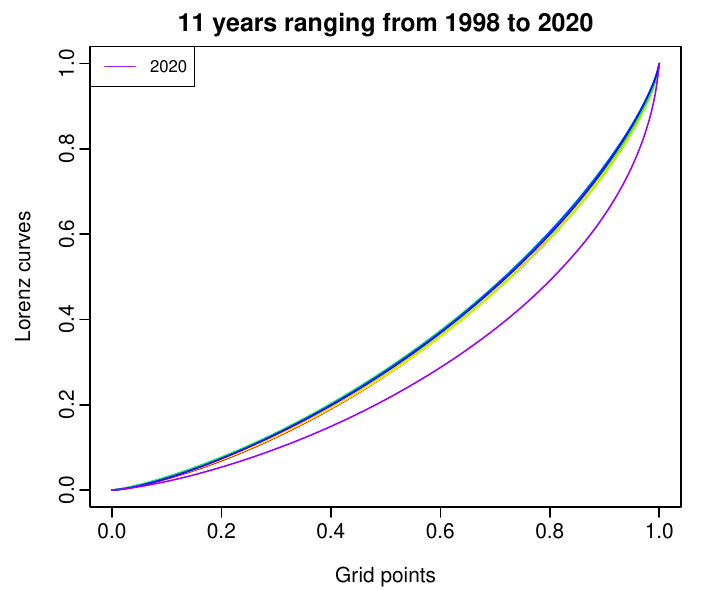}\label{fig:1a}}
\quad
\subfloat[Averaged logit transformation of Lorenz curves (spatial dimension) in 2020]
{\includegraphics[width=8.5cm]{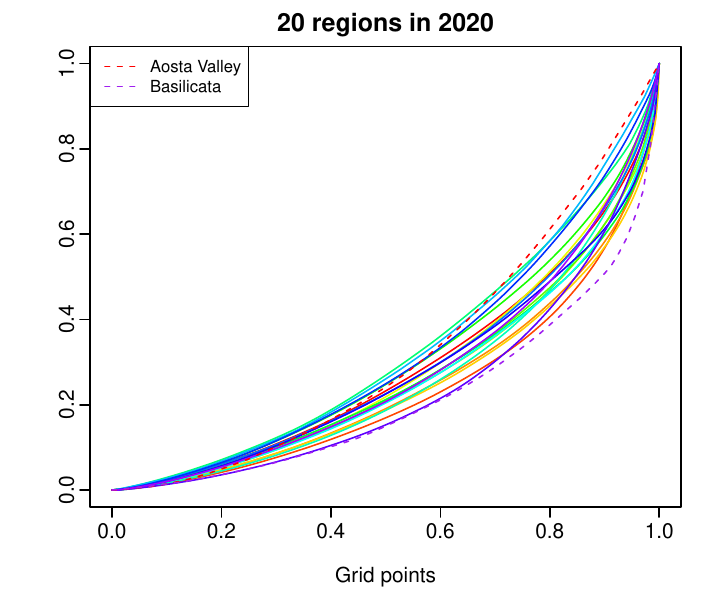}\label{fig:1b}}
\caption{\small The time series of the logit-transformed Lorenz curves observed from 1998 to 2020 averaged over 20 regions in Italy.}\label{fig:1}
\end{figure}

From Figure~\ref{fig:1a}, the Lorenz curves exhibit a similar shape, but there is an increase in income inequality in 2020, possibly due to the Covid-19 pandemic. For other years, the Lorenz curves were not equal, but the differences were quite small. The Lorenz curves behave like a set of CDFs, with values between 0 and 1. To remove the constraint of the restricted range, we implement a logit transformation \citep[see also][]{SH24}:
\begin{equation}
Y(u) = \ln \frac{L(u)}{1-L(u)},\quad u\in R.\label{eq:logit}
\end{equation} 
In 2020, Figure~\ref{fig:1b} displays the Lorenz curves for the 20 regions ordered geographically from North to South in Italy. Some variations exist from region to region, we can see greater income equality in a northern region (e.g., Aosta Valley) than in a southern region (e.g., Basilicata). \cite{Condino23} developed a classification tool to cluster the Lorenz curves of 20 regions into groups. In Table~\ref{tab:1}, we list the 20 regions in Italy ordered geographically from north to south.

\begin{table}[!htb]
\centering
\caption{\small The 20 regions in Italy ordered by geographical locations from north to south.}\label{tab:1}
\begin{tabular}{@{}ll@{}}
\toprule
Area 	& Region \\
\midrule
North 	& Piedmont, Aosta Valley, Lombardy, Trentino, Veneto, Friuli, Liguria, Emilia Romagna \\
Central 	& Tuscany, Umbria, Marche, Lazio, Abruzzo \\
South 	& Molise, Campania, Apulia, Basilicata, Calabria, Sicily, Sardinia \\
\bottomrule
\end{tabular}
\end{table}

\begin{figure}[!htb]
\centering
\includegraphics[width=7.3cm]{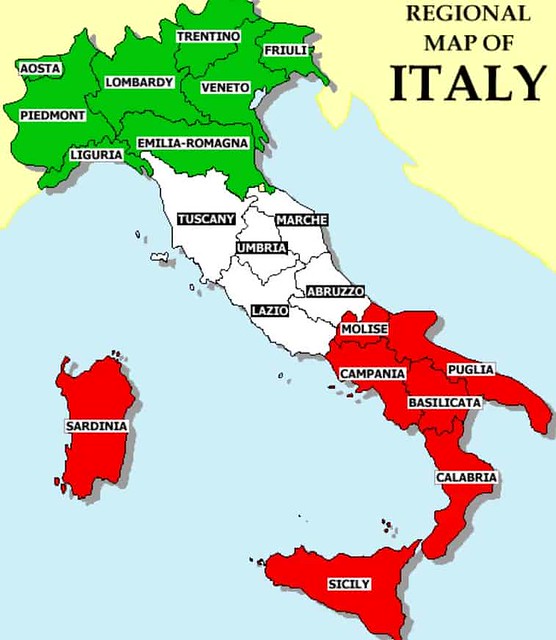}
\caption{\small A map of Italy showing 20 regions ordered geographically from North to South.}\label{fig:Italy_map}
\end{figure}

Figure~\ref{fig:Italy_map} displays a map of Italy showing 20 regions ordered geographically from North to~South.

\section{Forecasting high-dimensional functional time series}\label{sec:3}

The Lorenz curve is an example of a constrained functional time series, which has received increasing attention in the functional data analysis literature. Since the functional objects do not reside in a linear Hilbert space, there exist two schools of thought, namely intrinsic and extrinsic approaches. In the intrinsic approach, a distance metric, such as the Wasserstein metric, is used to measure distance. In the extrinsic approach, one-to-one transformations, such as log quantile transformation \citep{PM16}, centered log-ratio transformation \citep{KMP+19, SH20}, $\alpha$ transformation \citep{SH25, DSH+25}, and CDF transformation \citep{SH24}, are considered. We follow the extrinsic approach and adopt the CDF transformation.

Based on~\eqref{eq:logit}, we assume that random functions are sampled from a second-order stochastic process $Y$ in the square-integrable functions $\mathcal{L}^2$ residing in Hilbert space $\mathcal{H}$. Each realization $Y_t$ satisfies the condition $\|Y_t\|^2 = \int_{\mathcal{I}} Y^2_t(u)du<\infty$ with a function support range $\mathcal{I}\in R$. All random functions are defined on a common probability space with a finite second moment.

\subsection{Univariate functional time series forecasting}\label{sec:3.1}

For each region $s$, we implement a univariate functional time series forecasting method of \cite{HS09}. The method begins by computing an estimated covariance function, defined as
\begin{align*}
K^s(u,v) &= \text{Cov}[Y^s(u), Y^s(v)] \\
	      &= \text{E}\{[Y^s(u) - \mu^s(u)][Y^s(v) - \mu^s(v)]\},
\end{align*}
where $\mu^s(u)$ denotes the mean function at region $s$, where $s=1,\dots,20$ denotes each of the 20 regions in our data set. Since $K^s(u,v)$ is assumed to be a continuous and square-integrable covariance function, the function $K^s$ induces the kernel operator, given by
\begin{equation*}
(K^s\phi^s)(u) = \int_{\mathcal{I}} K^s(u,v)\phi^s(v)dv.
\end{equation*}
Assume that $K^s$ is continuous over $\mathcal{I}^2$, there exists an orthonormal sequence $(\phi^{s}_{k})$ of continuous function in $\mathcal{L}^2(\mathcal{I})$ and a non-increasing sequence $(\lambda^{s}_{k})$ of positive numbers, such that
\begin{equation*}
K^{s}(u,v) = \sum_{k=1}^{\infty}\lambda^{s}_{k}\phi^{s}_{k}(u)\phi^{s}_{k}(v), \quad u,v\in \mathcal{I}. 
\end{equation*}
With Mercer's lemma, the realizations of a stochastic process $Y^{s}_{t}(u)$ can be expressed as
\begin{align}
Y^{s}_{t}(u) &= \overline{Y}^s(u) + \sum_{k=1}^{\infty}\beta^{s}_{t,k}\phi^s_k(u) \notag\\
 		  &= \overline{Y}^s(u) + \sum_{k=1}^{K}\beta^{s}_{t,k}\phi^s_k(u)+e_{t}^{s}(u), \label{eq:1}
\end{align}
where $\overline{Y}^s(u) = \frac{1}{n}\sum^n_{t=1}Y^s_t(u)$ and $n$ denotes the number of years in the $s\textsuperscript{th}$ row, $\phi^s_k(u)$ represents $k\textsuperscript{th}$ estimated functional principal component for region $s$, $\beta_{t,k}^s$ denotes the $k\textsuperscript{th}$ estimated principal component scores for region $s$ and time $t$, $K$ denotes the number of retained components, and $e_t^s(u)$ denotes the residuals. Based on a set of residual functions $[e_1^s(u),\dots,e_n^s(u)]$, several hypothesis tests, including the independent test of \cite{GK07} and stationarity test of \cite{HKR14}, have been developed as diagnostic checks to examine temporal dimension. For any two regions, we implicitly assume that $Y_t^s$ and $Y_t^{s^{'}}$ are pairwise independent for any two regions $s\neq s^{'}$.

The selection of $K$ has received lots of attention in econometrics and statistics; some commonly adapted approaches include
\begin{inparaenum}
\item[1)] scree plots or the fraction of variance explained by the first few functional principal components \citep{Chiou12};
\item[2)] Akaike information criterion \citep{ANH15} and Bayesian information criterion \citep{OS24};
\item[3)] predictive cross validation leaving out one or more curves \citep{RS91};
\item[4)] bootstrap methods \citep{HV06}; and
\item[5)] eigenvalue ratio criterion \citep{AH13}. 
\end{inparaenum}

Following \cite{LRS20}, the value of $K^s$ is determined as the integer minimizing ratio of two adjacent empirical eigenvalues given by
\begin{equation*}
\widehat{K}^s = \argmin_{1\leq k\leq k_{max}}\left\{\frac{\widehat{\lambda}_{k+1}^s}{\widehat{\lambda}_k^s}\times \mathds{1}\Big(\frac{\widehat{\lambda}_k^s}{\widehat{\lambda}_1^s}\geq \delta\Big) + \mathds{1}\Big(\frac{\widehat{\lambda}_k^s}{\widehat{\lambda}_1^s}<\delta\Big)\right\},
\end{equation*}
where $k_{\max}$ is a pre-specified positive integer, $\delta$ is a pre-specified small positive number to trim off the smaller eigenvalues, and $\mathds{1}(\cdot)$ is the binary indicator function. We choose $k_{\max} = \#\{k|\widehat{\lambda}_k\geq \sum^n_{k=1}\widehat{\lambda}_k/n, k\geq 1\}$ and set the threshold constant $\delta = 1/\ln (\max(\widehat{\lambda}_1^s, n))$.

For a time series of functions $\bm{Y}^s(u) = [Y^s_1(u),\dots, Y^s_n(u)]$, we perform the FPCA to obtain the estimates of functional principal components $\bm{\Phi}^s(u)=[\phi_1^s(u),\dots,\phi_K^s(u)]$ and their associated scores $\bm{\widehat{\beta}}_k^s = [\widehat{\beta}_{1,k}^s,\dots,\widehat{\beta}_{n,k}^s]$. For each $k$, we apply a univariate time series forecasting method to $\bm{\widehat{\beta}}_k^s$ to obtain $\widehat{\beta}_{n+h|n,k}^s$, where $h$ denotes the forecast horizon. From~\eqref{eq:1}, the forecast curves can be obtained as
\begin{equation*}
\widehat{Y}_{n+h|n}^s(u) = \text{E}\big[Y_{n+h}^s(u)|\bm{Y}^s(u), \bm{\Phi}^s(u)\big] = \overline{Y}^s(u) + \sum_{k=1}^K\widehat{\beta}_{n+h|n,k}^s\phi_k^s(u).
\end{equation*}
Among the univariate time series forecasting methods, we consider the autoregressive integrated moving average (ARIMA) model. The order of ARIMA can be selected by an automatic algorithm of \cite{HK08} to choose the optimal orders of autoregressive~$p$, moving average~$q$, and difference order~$d$. The value of $d$ was selected based on successive Kwiatkowski-Phillips-Schmidt-Shin unit root tests. We applied the KPSS test to the original data; if the test result was significant, then we tested the differenced data for a unit root. The procedure terminates until we obtain our first insignificant result. Having determined $d$, the orders of $p$ and $q$ were selected based on the corrected Akaike information criterion.

By taking the inverse logit transformation, we obtain a $h$-step-ahead forecast of the Lorenz curve:
\begin{equation*}
\widehat{L}_{n+h|n}(p) = \frac{\exp[\widehat{Y}^s_{n+h|n}(u)]}{1+\exp[\widehat{Y}^s_{n+h|n}(u)]}.
\end{equation*}

\subsection{Construction of pointwise prediction intervals}\label{sec:3.2}

For measuring forecast uncertainty, prediction intervals based on statistical theory and data on error distributions provide an explicit estimate of the probability that future realizations lie within a given range. As studied in \cite{HS09}, the primary sources of uncertainty stem from 
\begin{inparaenum}
\item[(1)] the error in forecasting principal component scores;
\item[(2)] the model residuals.
\end{inparaenum}

Based on a univariate time series model, we can obtain forecasts for the principal component scores. Let $h$-step-ahead forecast errors be given by
\begin{equation*}
\nu^{s}_{\omega, k, h} = \widehat{\beta}^{s}_{\omega, k} - \widehat{\beta}^{s}_{\omega|\omega-h, k}, \quad k=1,2,\dots,K,
\end{equation*}
for $\omega=h+1,\dots,n$. These errors can be sampled with replacement to generate a bootstrap sample of $\beta_{n+h}$:
\begin{equation*}
\widehat{\beta}_{n+h|n, k}^{s, (b)} = \widehat{\beta}^s_{n+h|n, k}+\nu_{*, k, h}^{s, (b)}, \qquad b=1,\dots,B,
\end{equation*}
where $\nu_{*, k, h}^{s, (b)}$ are sampled with replacement from $\{\nu^s_{\omega, k, h}\}$, and $B=1,000$ represents the number of bootstrap samples.

When the functional principal component decomposition approximates the data well, the model residuals are random noise. Hence, we can bootstrap the model residuals in~\eqref{eq:1} by sampling with replacement from the model residual term $\{e_1^s(u),\dots, e_n^s(u)\}$.

Adding two sources of variability, we obtain $B$ variants for $Y_{n+h}^s(u)$,
\begin{equation}
\widehat{Y}_{n+h|n}^{s,(b)}(u) = \overline{Y}^s(u) + \sum_{k=1}^K\widehat{\beta}_{n+h|n, k}^{s,(b)}\phi_k^s(u) + e_{n+h}^{s,(b)}(u), \label{eq:3}
\end{equation}
where $\widehat{\beta}_{n+h|n, k}^{s,(b)}$ denotes the forecast of the bootstrapped principal component scores, and $e_{n+h}^{s,(b)}$ denotes the bootstrapped residual functions. With the bootstrapped $\left\{\widehat{Y}_{n+h|n}^{s,(1)}(u), \dots, \widehat{Y}_{n+h|n}^{s,(B)}(u)\right\}$, the pointwise prediction intervals are obtained by taking $\gamma/2$ and $1-\gamma/2$ quantiles at the $100(1-\gamma)\%$ nominal coverage probability. 

\subsection{One-way functional ANOVA}\label{sec:3.3}

Since we observe functional time series at each state, we are interested in examining the effect of the state, also known as the functional row effect. We resort to a decomposition known as one-way functional ANOVA \citep{Zhang13}. The observations can be decomposed as
\begin{equation}
Y^s_{t}(u) = \theta(u) + \eta^s(u) + \X^s_{t}(u), \label{eq:2}
\end{equation}
where $\theta(u)$ represents a functional grand effect, $\eta^{s}(u)$ denotes the $s\textsuperscript{th}$ functional row effect, and $\X^{s}_{t}(u)$ denotes the error term. To estimate $\eta^s(u)$ and $\X^{s}_{t}(u)$, we consider the functional median polish of \cite{SG12} because of its robustness. 

The functional median polish is an extension of the classic median polish by \cite{EH83}, and it is robust against outliers. Computationally, the functional grand effect and row effect can be extracted as
\begin{asparaenum}
\item[1)] Compute the functional median of each row and record the functional value. Subtract the row functional median from each function in that row.
\item[2)] Compute the functional median of the row functional medians, and record it as the functional grand effect. Subtract this functional grand effect from each of the row functional medians and record the values as the functional row effect. 
\item[3)] Repeat steps~1-2 and add the new functional grand effect and row effect to the current ones at each iteration until no changes occur with the row functional medians.
\end{asparaenum}
The above algorithm generally converges fast with one or two iterations, with constraints that $\text{median}_s\{\eta^s(u)\}=0$ and $\text{median}_s\{\X^s_{t}(u)\}=0$, $\forall t$. To compute the functional median, we use the modified band depth of \cite{LR09}, ranking curves from centre to outwards.

From~\eqref{eq:2}, we model $\X_{t}^s(u)$ by the univariate functional time series forecasting method in Section~\ref{sec:3.1}. With $\widehat{\X}_{n+h|n}^s(u)$, we add the deterministic part to obtain the $h$-step-ahead forecast curve as
\begin{equation*}
\widehat{Y}_{n+h|n}^s(u) = \theta(u) + \eta^s(u) + \widehat{\X}_{n+h|n}^s(u).
\end{equation*}

From~\eqref{eq:3}, we implement the nonparametric bootstrap method to simulate the $h$-step-ahead functional median polish residuals. By adding the deterministic part, the bootstrap forecast curves can be obtained as
\begin{equation*}
\widehat{Y}^{s, (b)}_{n+h|n}(u) = \theta(u) + \eta^s(u) + \widehat{\X}^{s, (b)}_{n+h|n}(u).
\end{equation*}
With the bootstrapped $\left\{\widehat{Y}_{n+h|n}^{s,(1)}(u), \dots, \widehat{Y}_{n+h|n}^{s,(B)}(u)\right\}$, the pointwise prediction intervals are obtained by taking $\gamma/2$ and $1-\gamma/2$ quantiles at the $100(1-\gamma)\%$ nominal coverage probability, where $\gamma$ represents a level of significance. 

We implement the isotonic regression to ensure the forecast CDF's monotonicity \citep[see also][]{Wied24}. In essence, among a set of grid points, the isotonic regression model locates the CDF values where the monotonic constraint does not satisfy and replaces them with their averages. Computationally, the isotonic regression can be carried out using the isoreg function in \Rlogo \ \citep{Team24}.

\section{Results}\label{sec:4}

\subsection{Illustration of functional median polish decomposition}\label{sec:4.1}

We apply the functional median polish to decompose region-specific Lorenz curves, which are HDFTS, into the functional grand effect, functional row effect, and residual functions. Since the Lorenz curves are constrained data, we consider a logit transformation to obtain transformed data. With a set of transformed data, the one-way functional ANOVA extracts the functional grand effect applied to all regions, while the functional row effect and functional residual are region-specific. 

\begin{figure}[!htb]
\centering
\subfloat[Functional grand effect]
{\includegraphics[width=6cm]{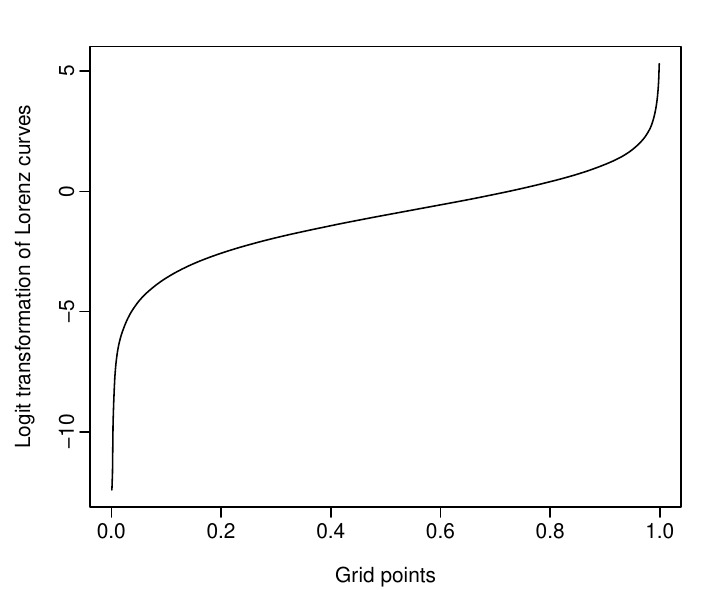}}
\subfloat[Functional row effect]
{\includegraphics[width=6cm]{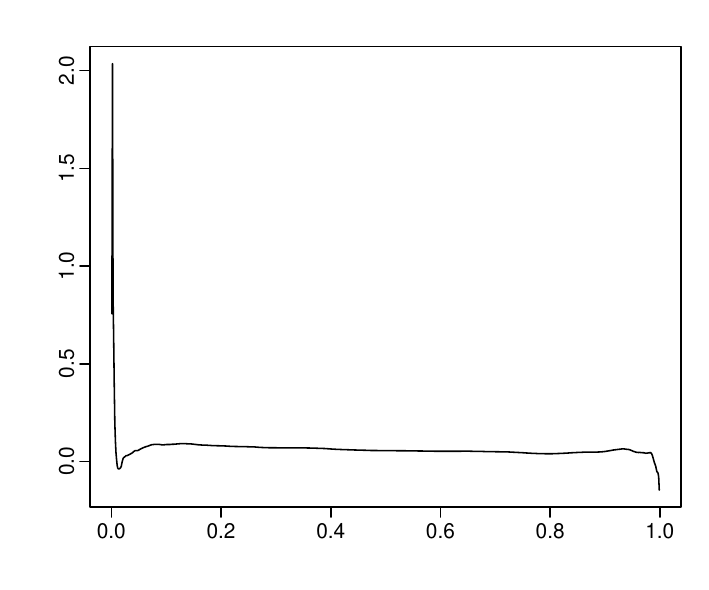}}
\subfloat[Residual functions]
{\includegraphics[width=6cm]{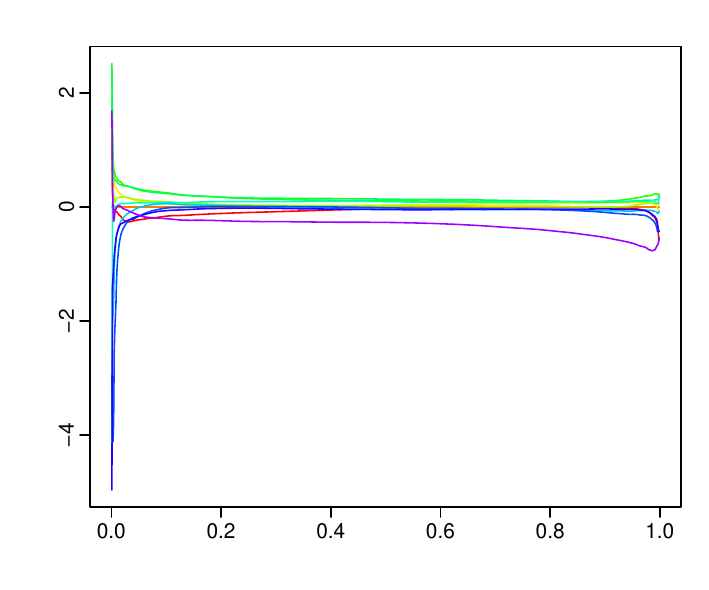}}
\\
\hspace{2.31in}
\subfloat[Curve reconstruction]
{\includegraphics[width=6cm]{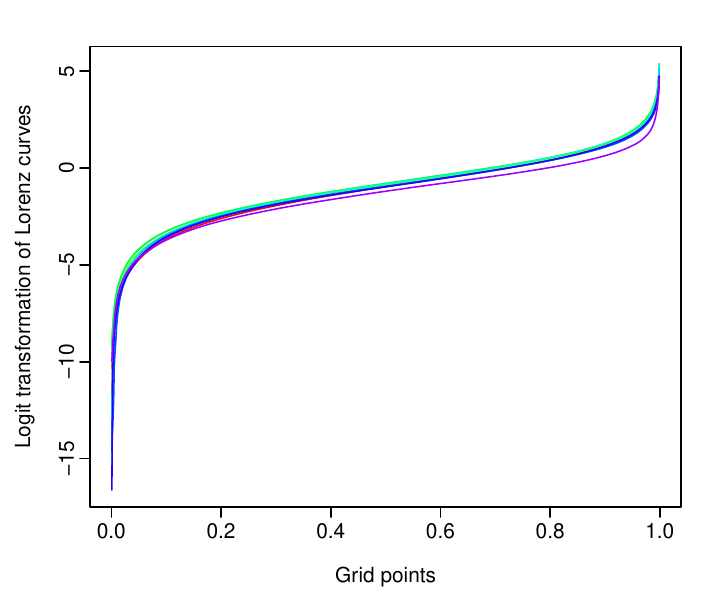}}
\subfloat[Original data]
{\includegraphics[width=6cm]{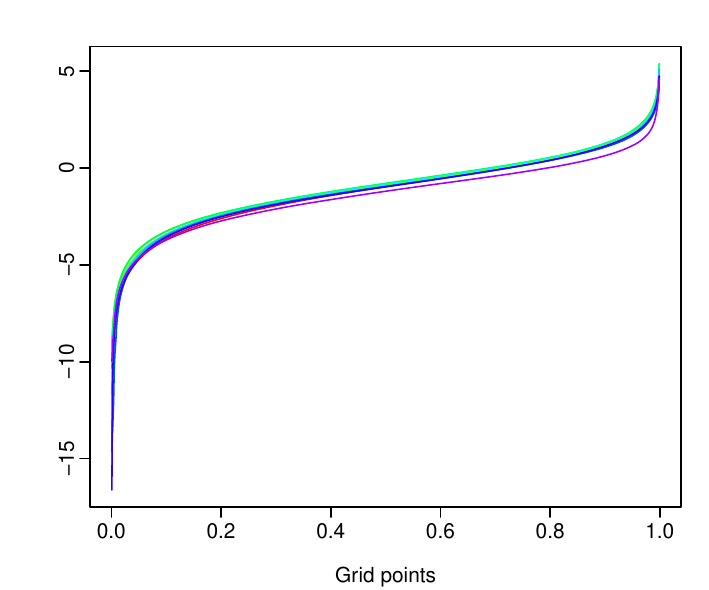}}
\caption{\small{One-way functional ANOVA decomposition of the logit transformed data for Piedmont.}}\label{fig:2}
\end{figure}

As a vehicle of illustration, we present the results for the first region (Piedmont). The three functions extracted from the functional median polish of Piedmont are displayed in the first row of Figure~\ref{fig:2}. By adding the three functions, the logit transformation of the Lorenz curves can be reconstructed and matched exactly with the original data. From a time series of the residual functions, we consider a stationarity test of \cite{HKR14}. From its $p$-value, we conclude that the residual functional time series is stationary.

\begin{figure}[!htb]
\centering
\subfloat[One-step-ahead forecast of the residual functions]
{\includegraphics[width=8cm]{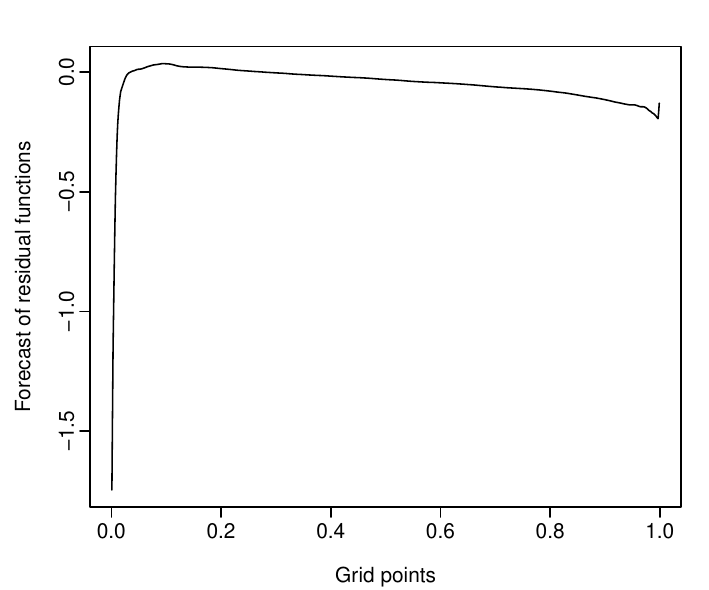}\label{fig:3a}}
\qquad
\subfloat[One-step-ahead forecast of the Lorenz curve]
{\includegraphics[width=8cm]{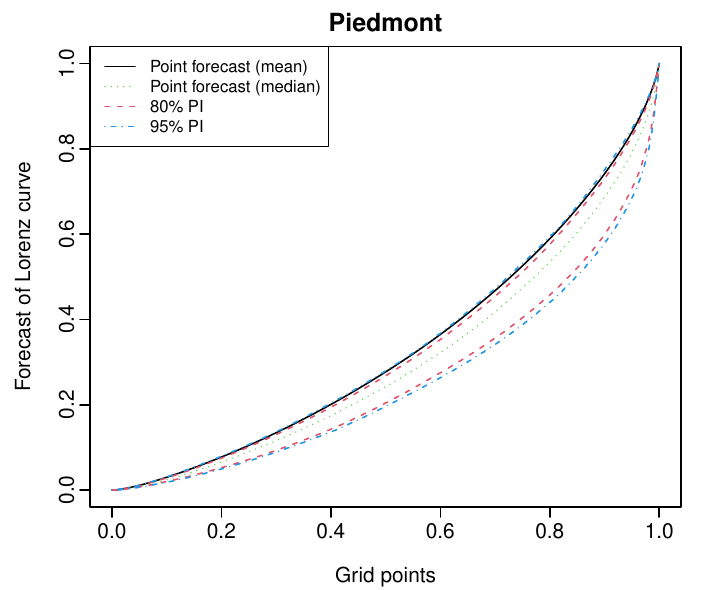}\label{fig:5}}
\caption{\small{One-step-ahead point and interval forecasts of the Lorenz curve for Piedmont.}}\label{fig:3}
\end{figure}

We implement a univariate functional time series forecasting method to forecast one-step-ahead residual functions in Figure~\ref{fig:3a}. By adding the functional grand effect and row effect in Figure~\ref{fig:2}, the one-step-ahead point and interval forecast curves can be obtained in Figure~\ref{fig:5}.


\subsection{One-step-ahead forecasts of Lorenz curves}\label{sec:4.2}

\begin{figure}[!htb]
\centering
\includegraphics[width=8.5cm]{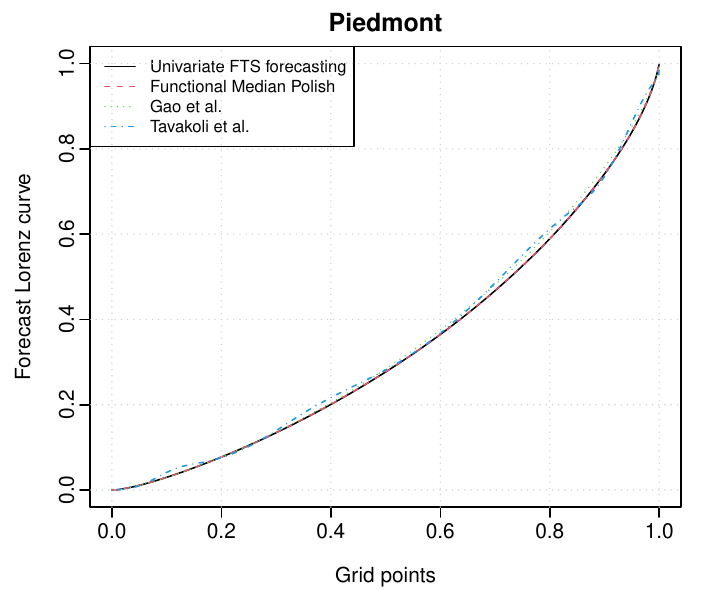}
\quad
\includegraphics[width=8.5cm]{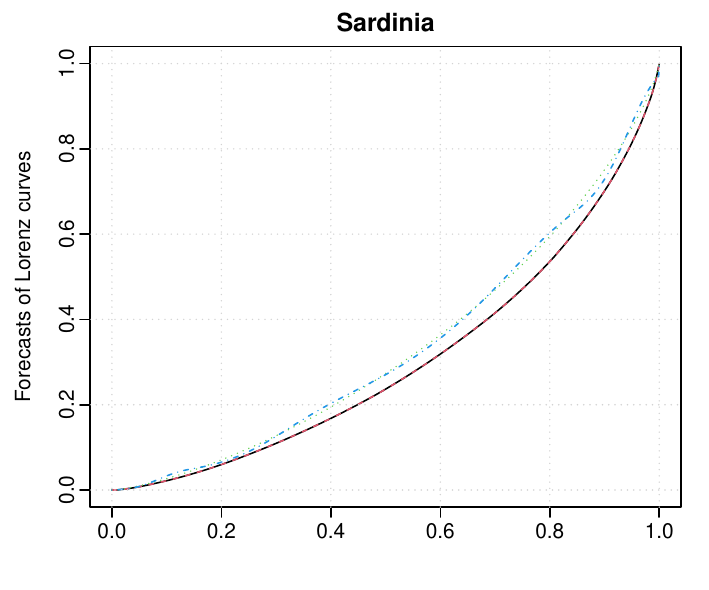}
\caption{\small One-step-ahead point forecasts of the Lorenz curves for the regions Piedmont and Sardinia in Italy.}\label{fig:4}
\end{figure}

Based on the historical data from 1998 to 2020, we implement the univariate functional time series forecasting and functional median polish methods to produce one-step-ahead forecast Lorenz curves at 20 Italian regions. Because of the similarity in curve shapes, we display the results for the regions Piedmont and Sardinia in Figure~\ref{fig:4}. The results for other regions can be obtained from the online supplement. For comparison, we include two functional factor models proposed by \cite{GSY19} and \cite{TNH23}, which are designed for modelling HDFTS. 

From Figure~\ref{fig:4}, the point forecasts are similar between the univariate functional time series forecasting method and functional median polish. This may be because the overall term seems to dominate the estimated curves and the residual functions are modelled via the same univariate functional time series forecasting method. Between the two regions, we can observe a difference in curve shape with a greater income equality in Piedmont than in Sardinia.

\subsection{Comparison of point forecast accuracy}

With 11 years of data, we split the data into the training and testing samples. The initial training sample consists of the data from 1998 to 2008, while the testing sample consists of the data from 2010 to 2020 with the exception of a missing year 2018. For forecasting, we consider the univariate functional time series forecasting method and the functional median polish method. For comparison, we also include the methods of \cite{GSY19} and \cite{TNH23}. The results obtained from \citeauthor{GSY19}'s \citeyearpar{GSY19} provide the most accurate point forecasts, as measured by the Kullback-Leibler (KL) divergence \citep{KL51} and the square root of the Jensen-Shannon divergence \citep{Shannon48}. 

The KL divergence is intended to measure the loss of information when we choose an approximation. For the actual and forecast Lorenz curves, denoted by $L^s_{m+\xi}(p)$ and $\widehat{L}^s_{m+\xi|m}(p)$, the discrete version of the KL divergence for a grid of 942 points is defined as
\begin{align*}
\text{KLD} =\ & D_{KL}[L^s_{m+\xi}(p_i)||\widehat{L}^s_{m+\xi|m}(p_i)]+D_{KL}[\widehat{L}^s_{m+\xi|m}(p_i)||L^s_{m+\xi}(p_i)] \\
=\ & \frac{1}{942\times 5}\sum_{\xi=1}^5\sum^{942}_{i=1}L^s_{m+\xi}(p_i)\cdot[\ln L^s_{m+\xi}(p_i) - \ln \widehat{L}^s_{m+\xi|m}(p_i)]+\\
&\frac{1}{942\times 5}\sum_{\xi=1}^5\sum^{942}_{i=1}\widehat{L}^s_{m+\xi|m}(p_i)\cdot[\ln \widehat{L}^s_{m+\xi|m}(p_i) - \ln L^s_{m+\xi}(p_i)],
\end{align*}
which is symmetric and non-negative. An alternative is given by the Jensen-Shannon divergence defined by
\begin{equation*}
    \text{JSD} =  \frac{1}{2}\text{D}_{\text{KL}}\left[L^s_{m+\xi}(p_i)||\delta^s_{m+\xi}(p_i) \right] +\frac{1}{2}\text{D}_{\text{KL}}[\widehat{L}^s_{m+\xi|m}(p_i)||\delta^s_{m+\xi}(p_i)],
\end{equation*}
where $\delta_{m+\xi}(p_i)$ measures a common quantity between $L^s_{m+\xi}(p_i)$ and $\widehat{L}^s_{m+\xi|m}(p_i)$. We consider geometric mean given by  $\delta^s_{m+\xi}(p_i)=\sqrt{L^s_{m+\xi}(p_i) \widehat{L}^s_{m+\xi|m}(p_i)}$. 

Table~\ref{tab:2} presents the KLD among the four methods. For one step ahead, the method of \cite{GSY19} provides the most accurate forecasts. The functional median polish method is notable for its ability to extract the functional grand and row effects, facilitating easier interpretation. Using the default tuning parameters in \citeauthor{TNH23}'s \citeyearpar{TNH23}, it produces inferior results for this data set. 

\begin{center}
\tabcolsep 0.133in
\begin{longtable}{@{}lllllllll@{}}
\caption{\small The KLD $(\times 100)$ and JSD $(\times 100)$ between the forecast and holdout Lorenz curves.} \label{tab:2} \\\hline
\toprule
  & \multicolumn{2}{c}{Univariate FTS} & \multicolumn{2}{c}{FMP} & \multicolumn{2}{c}{Gao et al.} & \multicolumn{2}{c}{Tavakoli et al.} \\ 
Region & KLD & JSD  & KLD & JSD  & KLD & JSD   & KLD & JSD   \\
\midrule
\endfirsthead
\toprule
  & \multicolumn{2}{c}{Univariate FTS} & \multicolumn{2}{c}{FMP} & \multicolumn{2}{c}{Gao et al.} & \multicolumn{2}{c}{Tavakoli et al.} \\ 
Region & KLD & JSD  & KLD & JSD  & KLD & JSD   & KLD & JSD   \\
\midrule
\endhead
\hline \multicolumn{9}{r}{{Continued on next page}} \\
\endfoot
\endlastfoot
  Piedmont 		& 0.0467 & 0.0117 & 0.0456 & 0.0114 & 0.0548 & 0.0137 & 0.1075 & 0.0271 \\ 
  Aosta Valley 		& 0.2595 & 0.0651 & 0.2609 & 0.0655 & 0.2074 & 0.0521 & 0.1986 & 0.0498 \\ 
  Lombardy 		& 0.1887 & 0.0472 & 0.1659 & 0.0415 & 0.1686 & 0.0421 & 0.1772 & 0.0444 \\ 
  Trentino 			& 0.1231 & 0.0308 & 0.1225 & 0.0307 & 0.0988 & 0.0247 & 0.1692 & 0.0424 \\ 
  Veneto 			& 0.0696 & 0.0174 & 0.0717 & 0.0179 & 0.0650 & 0.0163 & 0.1179 & 0.0295 \\ 
  Friuli 			& 0.0757 & 0.0190 & 0.0769 & 0.0193 & 0.0793 & 0.0199 & 0.1016 & 0.0256 \\ 
  Liguria 			& 0.0889 & 0.0222 & 0.0775 & 0.0194 & 0.0631 & 0.0158 & 0.1139 & 0.0285 \\ 
  Emilia Romagna 	& 0.0523 & 0.0131 & 0.0522 & 0.0131 & 0.0513 & 0.0128 & 0.0925 & 0.0232 \\ 
  Tuscany 			& 0.0212 & 0.0053 & 0.0216 & 0.0054 & 0.0318 & 0.0080 & 0.0605 & 0.0152 \\ 
  Umbria 			& 0.0798 & 0.0200 & 0.0779 & 0.0195 & 0.0849 & 0.0212 & 0.1190 & 0.0298 \\ 
  Marche 			& 0.0116 & 0.0029 & 0.0120 & 0.0030 & 0.0161 & 0.0040 & 0.0500 & 0.0125 \\ 
  Lazio 			& 0.1621 & 0.0406 & 0.1797 & 0.0450 & 0.1769 & 0.0443 & 0.2281 & 0.0571 \\ 
  Abruzzo 			& 0.0882 & 0.0221 & 0.0884 & 0.0222 & 0.0640 & 0.0160 & 0.0993 & 0.0250 \\ 
  Molise 			& 0.0250 & 0.0065 & 0.0270 & 0.0070 & 0.0260 & 0.0066 & 0.1033 & 0.0262 \\ 
  Campania 		& 0.0246 & 0.0062 & 0.0215 & 0.0054 & 0.0284 & 0.0071 & 0.0857 & 0.0215 \\ 
  Apulia 			& 0.0200 & 0.0050 & 0.0187 & 0.0047 & 0.0310 & 0.0078 & 0.0710 & 0.0178 \\ 
  Basilicata 		& 0.2928 & 0.0732 & 0.2932 & 0.0733 & 0.2499 & 0.0625 & 0.2995 & 0.0750 \\ 
  Calabria 			& 0.0247 & 0.0062 & 0.0246 & 0.0062 & 0.0253 & 0.0064 & 0.0749 & 0.0189 \\ 
  Sicily 			& 0.3625 & 0.0906 & 0.3637 & 0.0909 & 0.2703 & 0.0675 & 0.4242 & 0.1060 \\ 
  Sardinia 			& 0.0978 & 0.0245 & 0.0978 & 0.0245 & 0.0991 & 0.0248 & 0.1794 & 0.0449 \\ 
  \midrule
  Mean 			& 0.1057 & 0.0265 & 0.1050 & 0.0263 & \textBF{0.0946} & \textBF{0.0237} & 0.1437 & 0.0360 \\
\bottomrule  
\end{longtable}
\end{center}

\vspace{-.5in}

\begin{figure}[!htb]
\centering
{\includegraphics[width=8.35cm]{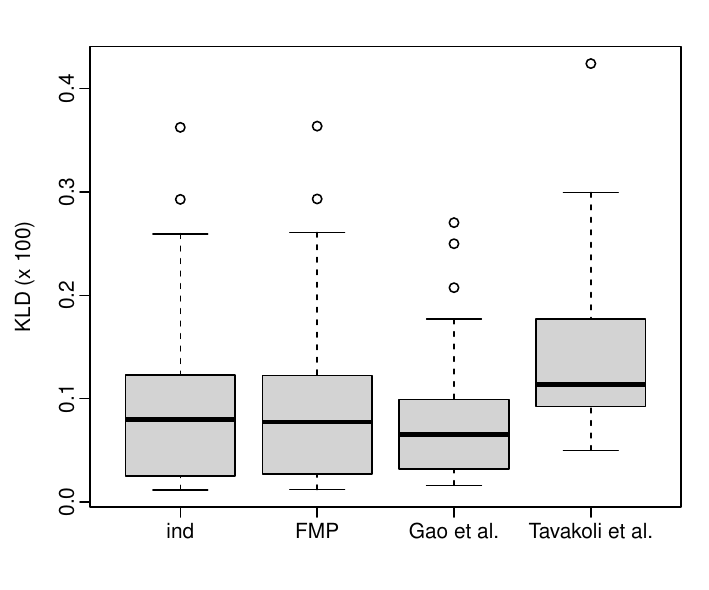}}
\qquad
{\includegraphics[width=8.35cm]{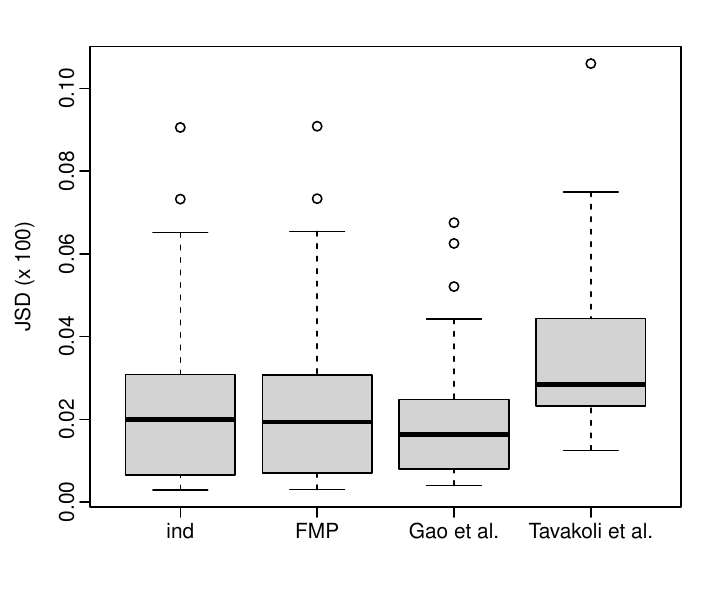}}
\caption{\small Boxplots of the one-step-ahead forecast errors, measured by KLD and JSD, among the four methods.}\label{fig:6}
\end{figure}

To visualize the overall results of the point forecast accuracy, we display the KLD and JSD using boxplots in Figure~\ref{fig:6}.

\subsection{Comparison of interval forecast accuracy}\label{sec:}

To evaluate and compare the interval forecast accuracy, we consider the interval score of \cite{GR07} and \cite{GK14}. For each observation in the forecasting period, the one-step-ahead prediction intervals are computed at the $100(1-\alpha)\%$ nominal coverage probability, where $\alpha$ denotes a significance level. We consider the common case of the symmetric $100(1-\alpha)\%$ prediction intervals, with lower and upper bounds that are predictive quantiles, denoted by $\widehat{L}^{s,\text{lb}}_{m+\xi|m}(p_i)$ and $\widehat{L}^{s,\text{ub}}_{m+\xi|m}(p_i)$. We compute the empirical coverage probability (ECP), defined as
\[
\text{ECP} = 1 - \frac{1}{942 \times 5}\sum_{\xi=1}^5\sum_{i=1}^{942}\left[\mathds{1}\{L^s_{m+\xi}(p_i)>\widehat{L}^{s,\text{ub}}_{m+\xi|m}(p_i)\}+\mathds{1}\{L^s_{m+\xi}(p_i)<\widehat{L}^{s,\text{lb}}_{m+\xi|m}(p_i)\}\right],
\]
where $\mathds{1}\{\cdot\}$ represents the binary indicator function, and $m=6$ denotes the length of the initial training sample. While the empirical coverage probability reveals over- or under-estimation of the nominal coverage probability, it is not an accuracy criterion due to the possible cancellation. As an alternative, the CPD is defined as
\[
\text{CPD} = \left|\text{ECP} - (1-\alpha)\right|.
\]
The smaller the value of CPD is, the better the method is. Although the empirical coverage probability and CPD are measures of interval forecast accuracy, neither they consider the sharpness of the prediction intervals, i.e., the distance between the lower and upper bounds. To rectify the problem, as defined by \cite{GR07}, a scoring rule for the interval forecasts at time point $\X_{m+\xi}(u_i)$ is
\begin{align*}
S_{\alpha,\xi}\big[\widehat{L}_{m+\xi|m}^{s,\text{lb}}(p_i), \widehat{L}_{m+\xi|m}^{s,\text{ub}}(p_i), & L^s_{m+\xi}(p_i)\big] = \left[\widehat{L}_{m+\xi|m}^{s,\text{ub}}(p_i) - \widehat{L}_{m+\xi|m}^{s,\text{lb}}(p_i)\right] \\
&+\frac{2}{\alpha}\left[\widehat{L}_{m+\xi|m}^{s,\text{lb}}(p_i)-L^s_{m+\xi}(p_i)\right]\mathds{1}\left\{L^s_{m+\xi}(p_i)<\widehat{L}_{m+\xi}^{s,\text{lb}}(p_i)\right\} \\
&+\frac{2}{\alpha}\left[L^s_{m+\xi}(p_i) - \widehat{L}_{m+\xi|m}^{s,\text{ub}}(p_i)\right]\mathds{1}\left\{L^s_{m+\xi}(p_i)>\widehat{L}_{m+\xi}^{s,\text{ub}}(p_i)\right\}.
\end{align*}
The interval score rewards a narrow prediction interval if and only if the holdout observations lie between the prediction interval. The optimal interval score is achieved when $L^s_{m+\xi}(p_i)$ lies between $\widehat{L}_{m+\xi}^{s,\text{lb}}(p_i)$ and $\widehat{L}_{m+\xi}^{s,\text{ub}}(p_i)$ in a frequency close to the nominal coverage probability, and the pointwise distance between $\widehat{L}_{m+\xi}^{s,\text{ub}}(p_i)$ and $\widehat{L}_{m+\xi}^{s,\text{lb}}(p_i)$ is minimal.

Table~\ref{tab:3} presents one-step-ahead interval forecast errors, as measured by the mean interval scores and CPD. The method of \cite{TNH23} does not produce the interval forecasts. Hence, we evaluate and compare the interval forecast accuracy among the univariate functional time series method, one-way functional median polish, and the method of \cite{GSY19}. The one-way functional median polish and the method of \cite{TNH23} provide a smaller mean interval score. Based on the CPD, the one-way functional median polish produces the most accurate interval forecasts.

\begin{center}
\tabcolsep 0.075in
\begin{longtable}{@{}llrrrrrr@{}}
\caption{\small For computing 80\% and 95\% prediction intervals, we compute one-step-ahead mean interval scores and CPD among the univariate functional time series method, one-way functional median polish, and the method of \cite{GSY19}.} \label{tab:3} \\\hline
\toprule
& & \multicolumn{3}{c}{Interval score} & \multicolumn{3}{c}{CPD} \\
$\alpha$ & Region & Univariate FTS & FMP & Gao et al. & Univariate FTS & FMP & Gao et al. \\ 
\midrule
\endfirsthead
\toprule
& & \multicolumn{3}{c}{Interval score} & \multicolumn{3}{c}{CPD} \\
$\alpha$ & Region & Univariate FTS & FMP & Gao et al. & Univariate FTS & FMP & Gao et al. \\ 
\midrule
\endhead
\hline \multicolumn{8}{r}{{Continued on next page}} \\
\endfoot
\endlastfoot
0.2 	& Piedmont 		& 0.1315 & 0.1117 & 0.1537 & 0.4129 & 0.3124 & 0.5214 \\ 
  	& Aosta Valley 		& 0.1286 & 0.1174 & 0.0998 & 0.3101 & 0.1931 & 0.2823 \\ 
	& Lombardy 		& 0.1702 & 0.1741 & 0.1908 & 0.2810 & 0.2928 & 0.4581 \\ 
  	& Trentino 		& 0.2315 & 0.2232 & 0.1979 & 0.5607 & 0.5775 & 0.4590 \\ 
  	& Veneto 			& 0.2189 & 0.1707 & 0.1636 & 0.5257 & 0.2924 & 0.3751 \\ 
  	& Friuli 			& 0.1213 & 0.1066 & 0.1076 & 0.2740 & 0.2806 & 0.3369 \\ 
  	& Liguria 			& 0.2002 & 0.1727 & 0.1395 & 0.6208 & 0.4037 & 0.3344 \\ 
  	& Emilia Romagna 	& 0.1568 & 0.1366 & 0.1332 & 0.3490 & 0.2646 & 0.2731 \\ 
  	& Tuscany 		& 0.1118 & 0.0764 & 0.1142 & 0.3899 & 0.2927 & 0.3487 \\ 
  	& Umbria 			& 0.2012 & 0.2043 & 0.2126 & 0.2568 & 0.2580 & 0.3027 \\ 
  	& Marche 			& 0.0865 & 0.0470 & 0.0986 & 0.3466 & 0.2179 & 0.3953 \\ 
  	& Lazio 			& 0.1360 & 0.1453 & 0.1405 & 0.3901 & 0.3009 & 0.2959 \\ 
  	& Abruzzo 		& 0.1730 & 0.1918 & 0.1575 & 0.3630 & 0.3818 & 0.2986 \\ 
  	& Molise 			& 0.0383 & 0.0289 & 0.0550 & 0.2124 & 0.1368 & 0.1843 \\ 
  	& Campania 		& 0.0869 & 0.0868 & 0.0933 & 0.3507 & 0.2999 & 0.3630 \\ 
  	& Apulia 			& 0.0438 & 0.0557 & 0.0522 & 0.2317 & 0.2378 & 0.2965 \\ 
  	& Basilicata 		& 0.2792 & 0.2741 & 0.2175 & 0.4237 & 0.4352 & 0.3628 \\ 
  	& Calabria 		& 0.1135 & 0.1207 & 0.1196 & 0.3837 & 0.3814 & 0.3782 \\ 
  	& Sicily 			& 0.1741 & 0.1810 & 0.1505 & 0.2971 & 0.3051 & 0.3905 \\ 
  	& Sardinia 		& 0.1343 & 0.1331 & 0.1236 & 0.3549 & 0.2398 & 0.2594 \\ 
\cmidrule{2-8}
  	& Mean 			& 0.1469 & 0.1379 & \textBF{0.1361} & 0.3667 & \textBF{0.3052} & 0.3458 \\ 
\midrule
0.05 &   Piedmont 		& 0.3833 & 0.2755 & 0.4481 & 0.4607 & 0.3388 & 0.4998 \\ 
	&   Aosta Valley 	& 0.2188 & 0.2026 & 0.1016 & 0.2878 & 0.1450 & 0.1284 \\ 
	&   Lombardy 		& 0.5753 & 0.5833 & 0.6563 & 0.2537 & 0.2215 & 0.4006 \\ 
	&   Trentino 		& 0.7125 & 0.6130 & 0.5877 & 0.5940 & 0.5891 & 0.2471 \\ 
	&   Veneto 		& 0.7352 & 0.5572 & 0.4611 & 0.5492 & 0.2187 & 0.2954 \\ 
	&   Friuli 			& 0.2873 & 0.2080 & 0.2659 & 0.2813 & 0.2456 & 0.3129 \\ 
	&   Liguria 		& 0.6489 & 0.5332 & 0.4171 & 0.5891 & 0.3542 & 0.3327 \\ 
	&   Emilia Romagna 	& 0.5448 & 0.4573 & 0.4831 & 0.3878 & 0.2071 & 0.2388 \\ 
	&   Tuscany 		& 0.3163 & 0.1724 & 0.2598 & 0.3622 & 0.2399 & 0.3826 \\ 
	&   Umbria 		& 0.6077 & 0.6112 & 0.6498 & 0.2289 & 0.2202 & 0.2480 \\ 
	&   Marche 		& 0.1964 & 0.0550 & 0.1574 & 0.3354 & 0.1028 & 0.3804 \\ 
	&   Lazio 			& 0.3991 & 0.4383 & 0.4788 & 0.2373 & 0.2253 & 0.2215 \\ 
	&   Abruzzo 		& 0.4715 & 0.5279 & 0.4466 & 0.3339 & 0.3251 & 0.2026 \\ 
	&   Molise 			& 0.0817 & 0.0386 & 0.0799 & 0.1847 & 0.0553 & 0.0992 \\ 
	&   Campania 		& 0.2775 & 0.2780 & 0.2986 & 0.3378 & 0.2030 & 0.3361 \\ 
	&   Apulia 			& 0.0608 & 0.0817 & 0.0573 & 0.1707 & 0.1283 & 0.1480 \\ 
	&   Basilicata 		& 0.7689 & 0.6901 & 0.5588 & 0.3522 & 0.3893 & 0.3361 \\ 
	&   Calabria 		& 0.2373 & 0.2452 & 0.2802 & 0.2925 & 0.3337 & 0.2700 \\ 
	&   Sicily 			& 0.5700 & 0.5862 & 0.4245 & 0.2243 & 0.2285 & 0.2198 \\ 
	&   Sardinia 		& 0.4646 & 0.4431 & 0.3972 & 0.3844 & 0.1987 & 0.2649 \\ 
\cmidrule{2-8}
	&   Mean 			& 0.4279 & 0.3799 & \textBF{0.3755} & 0.3424 & \textBF{0.2485} & 0.2782 \\   
\bottomrule  
\end{longtable}
\end{center}

\vspace{-.5in}

\section{Conclusion}\label{sec:5}

The Lorenz curve plays a vital role in economics for measuring income inequality at the national and regional levels. The regional Lorenz curves resemble similarities to a group of CDFs. We take the logit transformation to model unconstrained data via several high-dimensional functional time series methods. Among them, we consider the factor models of \cite{GSY19} and \cite{TNH23} and the one-way functional median polish method. The one-way functional median polish can robustly decompose a group of functional time series into a functional grand effect, a functional row effect and residual functions. By modelling the time-varying residual functions, we obtain one-step-ahead point and interval forecasts by a univariate functional time series method. After taking the inverse logit transformation, we obtain the one-step-ahead forecast curves by adding the forecast residual functions to the deterministic parts, including the functional grand and row effects.

We investigate the one-step-ahead point and interval forecast accuracies using the Italian household income and wealth data set from 1998 to 2020. As measured by the Kullback-Leibler and Jensen-Shannon divergences, the factor model of \cite{GSY19} provides the most accurate point forecasts. As measured by the mean interval scores and coverage probability difference, the one-way functional median polish presents the smallest interval forecast errors. 

Addressing income inequality across Italy's 20 regions demands policies that balance regional development with national cohesion. Economic measures should focus on investing in underdeveloped areas, particularly the less prosperous Southern regions, and providing tax incentives to stimulate regional growth.

There are at least three ways in which the presented methodology can be further extended:
\begin{inparaenum}
\item[1)] The Lorenz curves can further be disaggregated by other factors, such as socioeconomic status. In this case, one may consider two-way functional median polish in \cite{SG12} and \cite{JSS23}.
\item[2)] We study the one-step-ahead forecast accuracies. If the data series is longer, one can investigate multiple-step-ahead forecast accuracies.
\item[3)] We model the data using the functional time series forecasting method proposed by \cite{HS09}. However, alternative forecasting methods, such as the approach developed by \cite{HZ18}, can also be applied.
\end{inparaenum}

\newpage

\begin{center}
\large Supplementary material of ``Forecasting a time series of Lorenz curves: One-way functional analysis of variance"
\end{center}

\begin{figure}[!htb]
\centering
\includegraphics[width=4.4cm]{Fig_4_1}
\includegraphics[width=4.4cm]{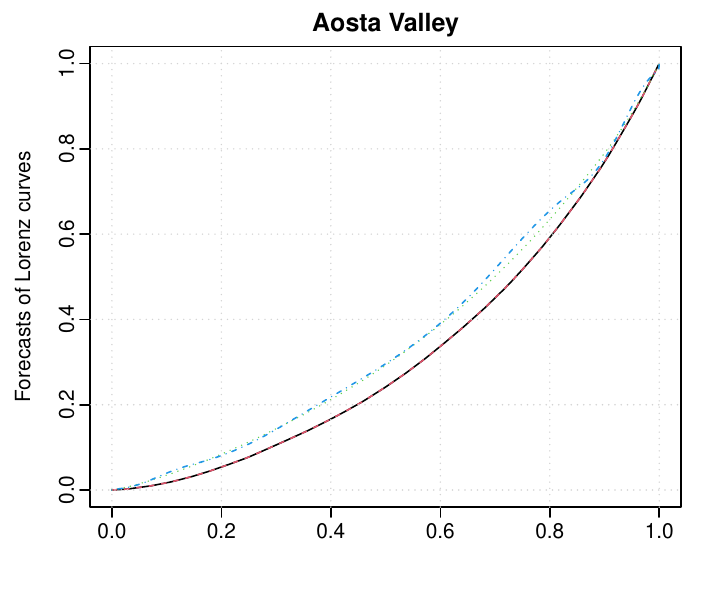}
\includegraphics[width=4.4cm]{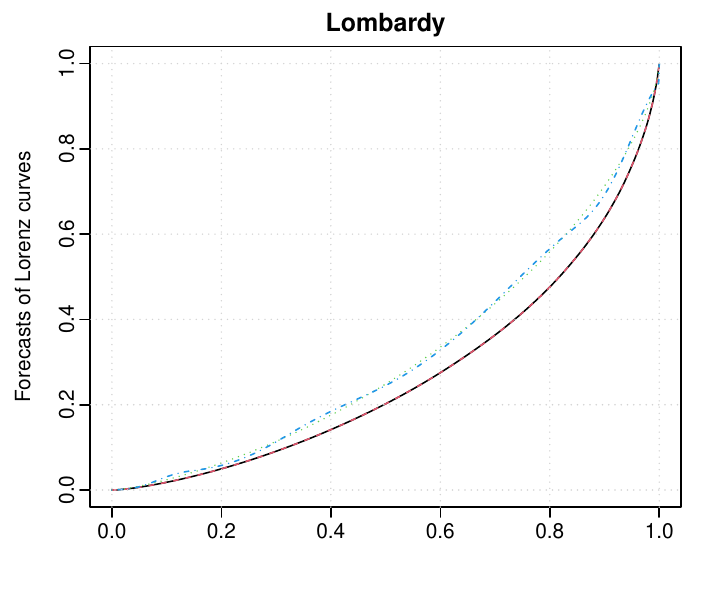}
\includegraphics[width=4.4cm]{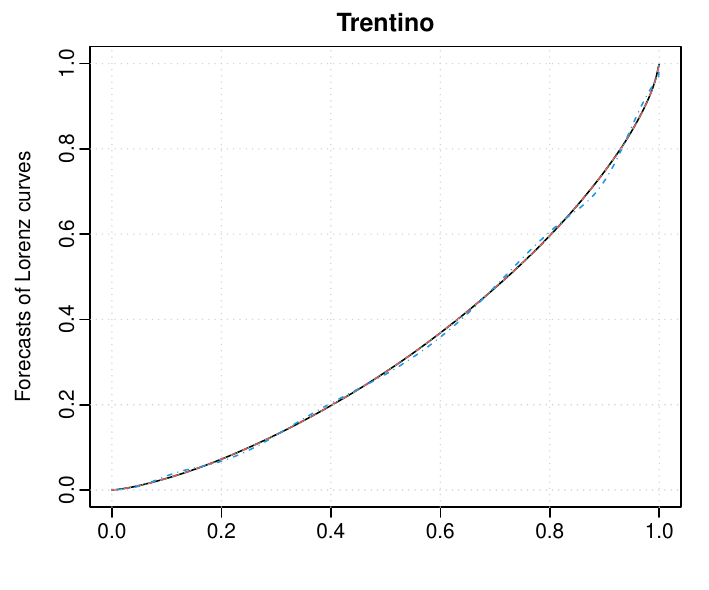}
\\
\includegraphics[width=4.4cm]{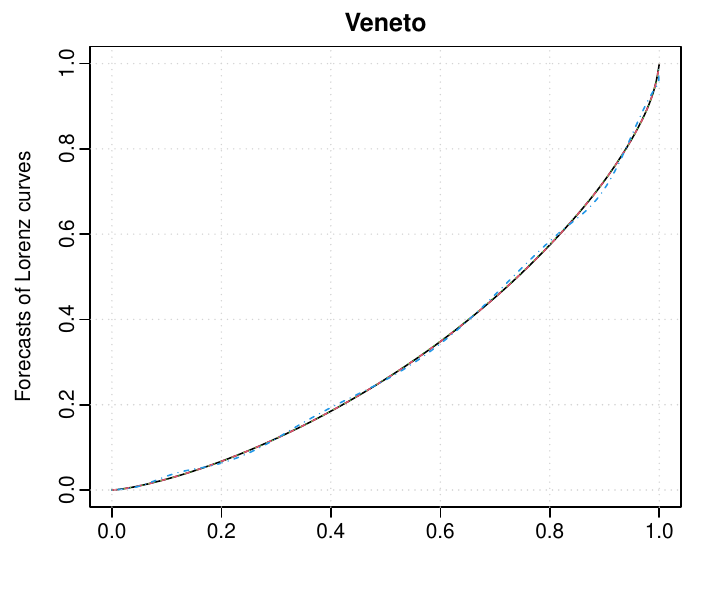}
\includegraphics[width=4.4cm]{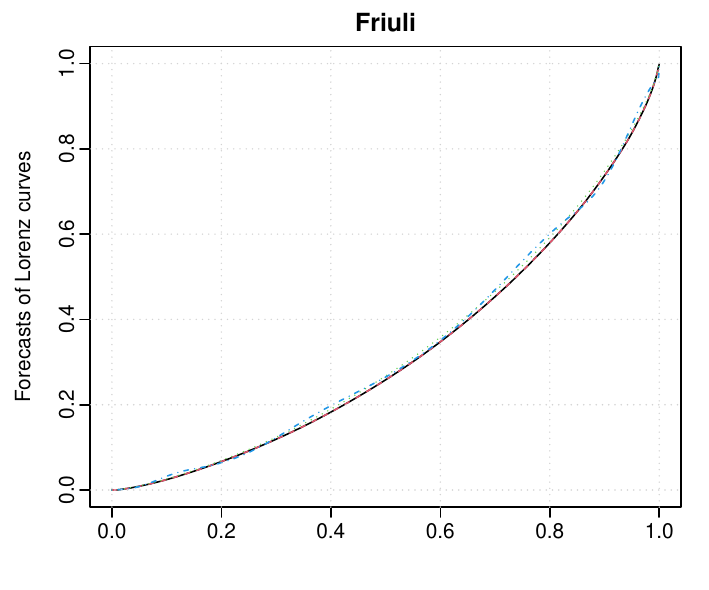}
\includegraphics[width=4.4cm]{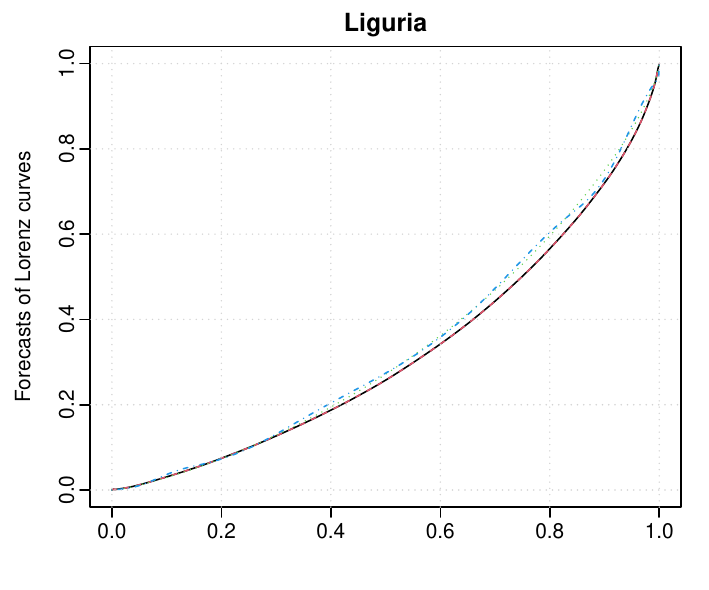}
\includegraphics[width=4.4cm]{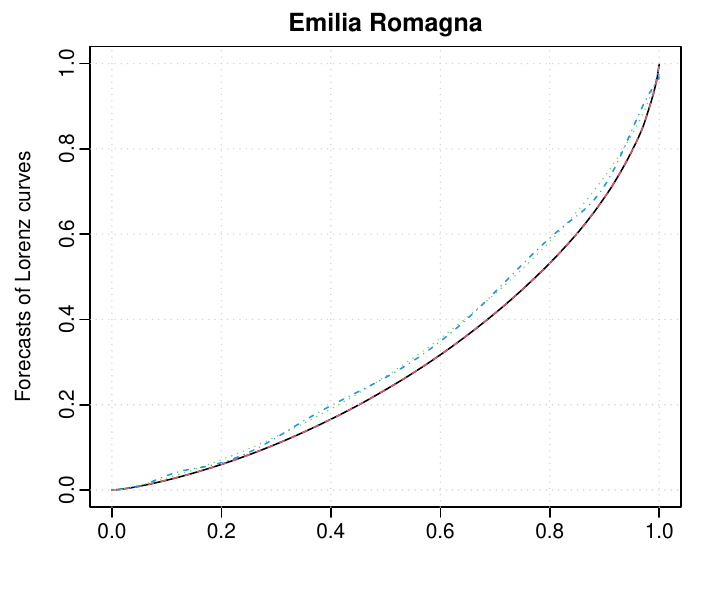}
\\
\includegraphics[width=4.4cm]{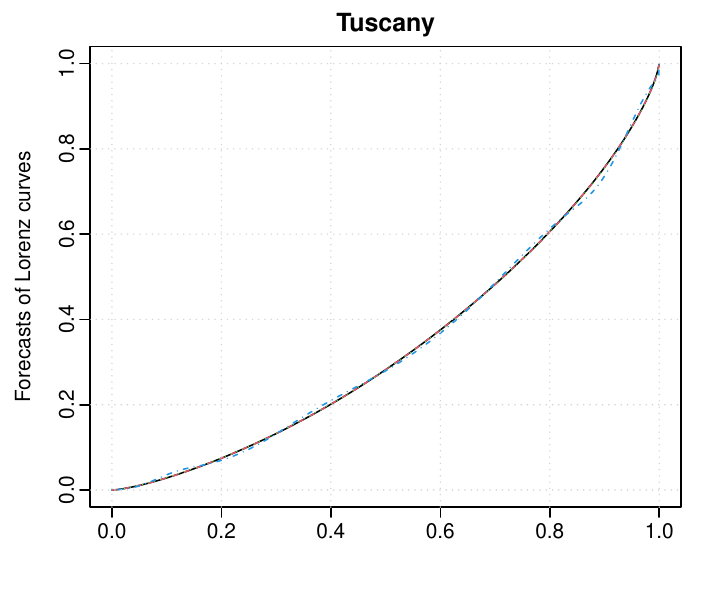}
\includegraphics[width=4.4cm]{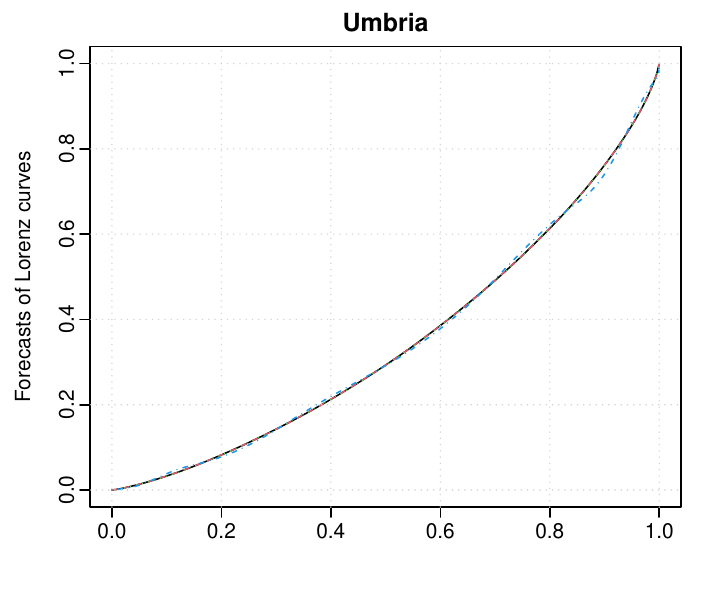}
\includegraphics[width=4.4cm]{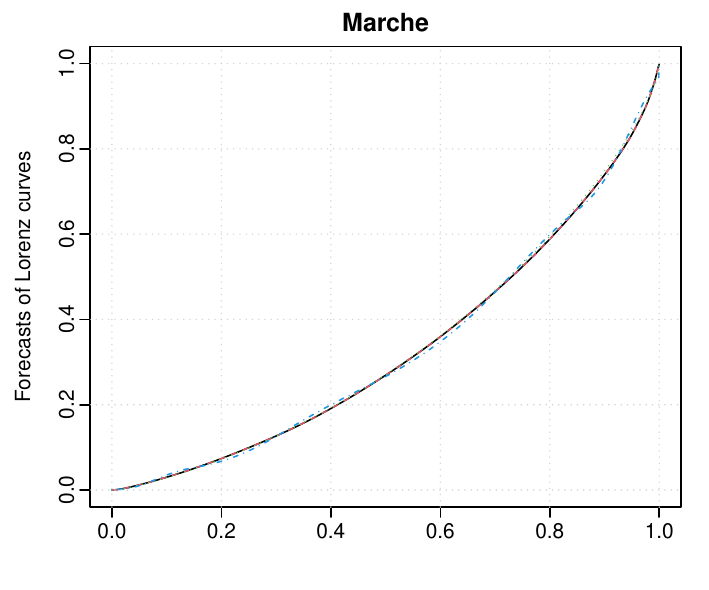}
\includegraphics[width=4.4cm]{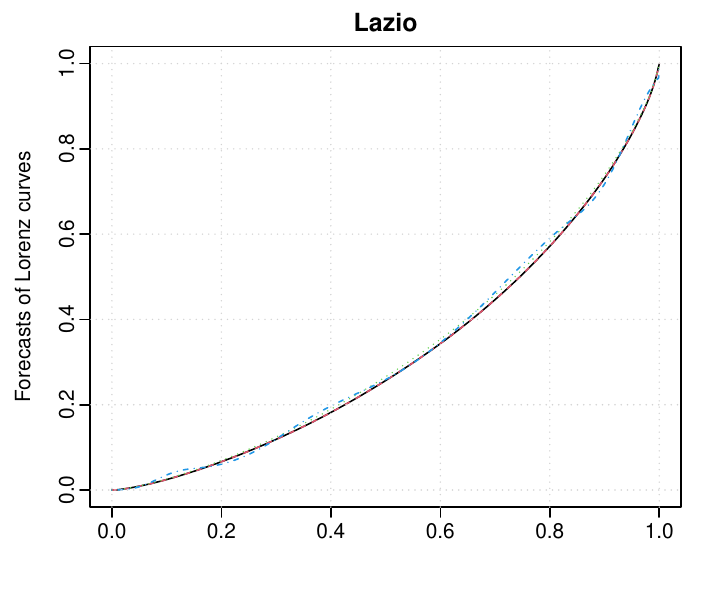}
\\
\includegraphics[width=4.4cm]{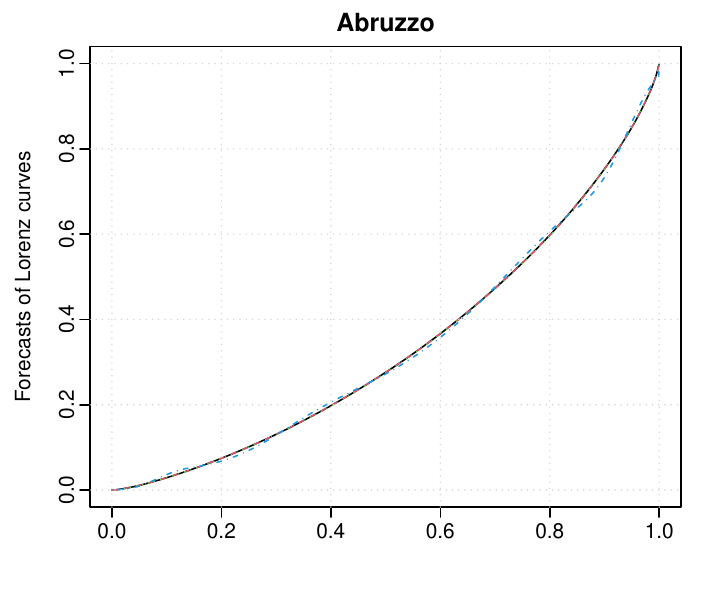}
\includegraphics[width=4.4cm]{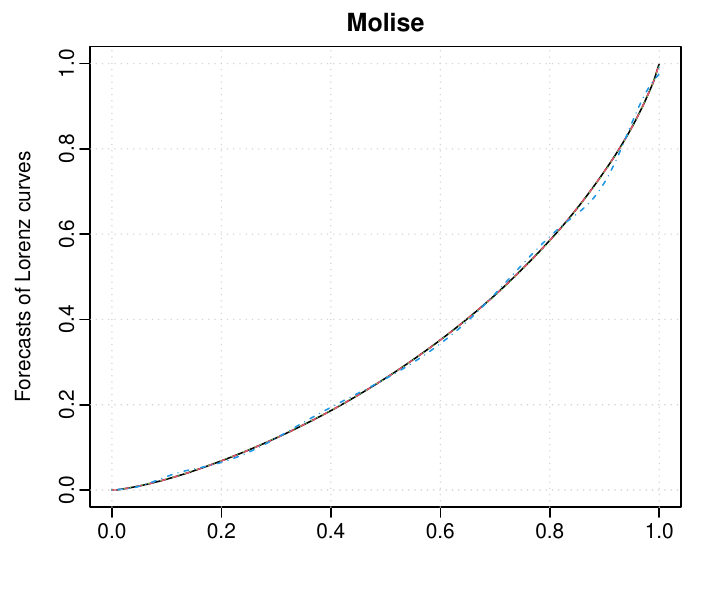}
\includegraphics[width=4.4cm]{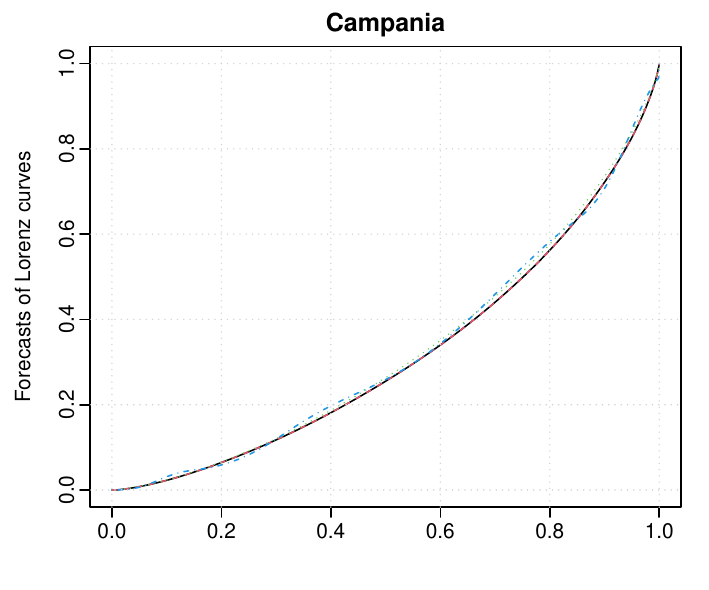}
\includegraphics[width=4.4cm]{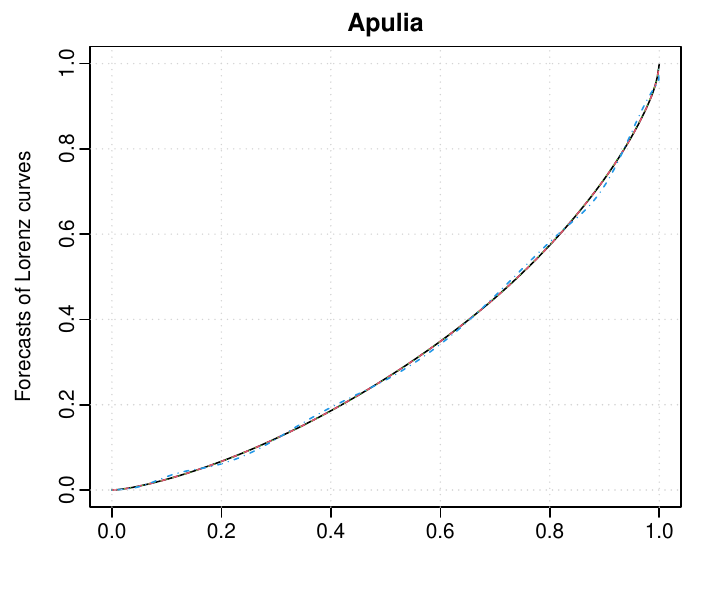}
\\
\includegraphics[width=4.4cm]{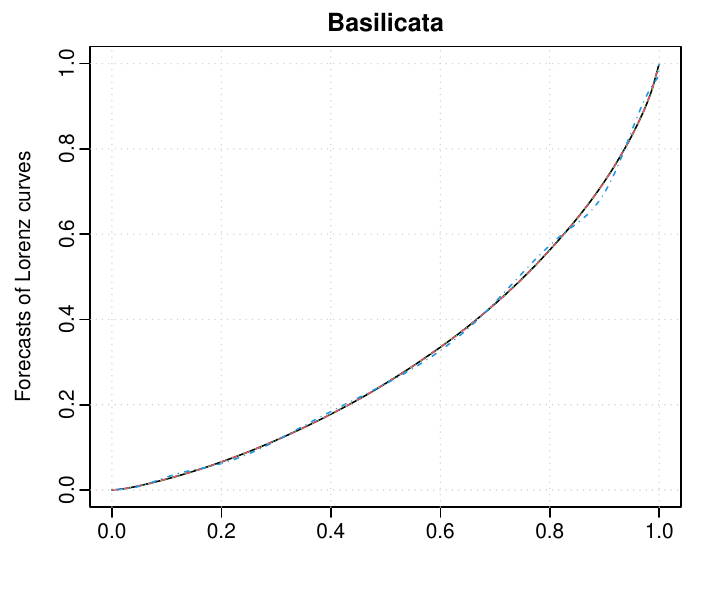}
\includegraphics[width=4.4cm]{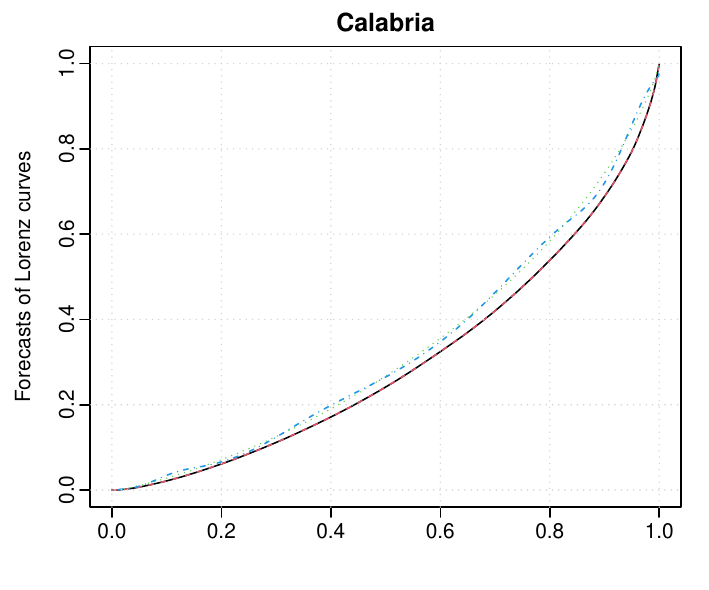}
\includegraphics[width=4.4cm]{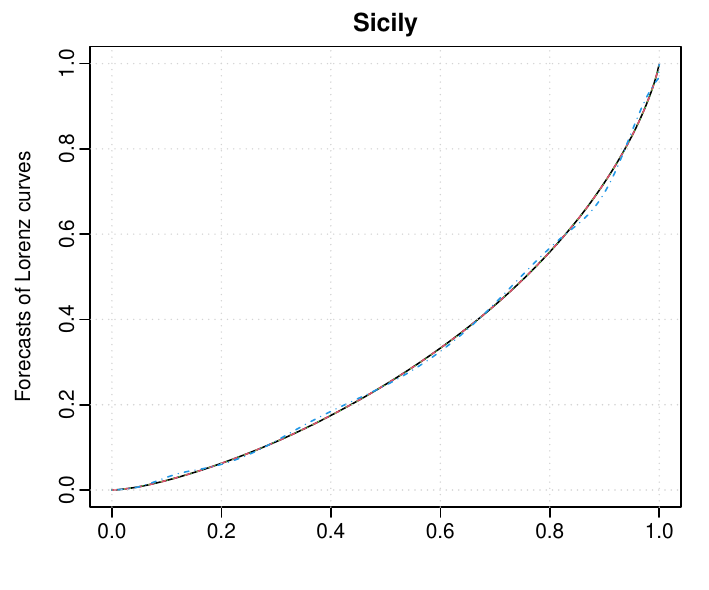}
\includegraphics[width=4.4cm]{Fig_4_20}
\caption{\small One-step-ahead point forecasts of the Lorenz curves for the 20 regions in Italy.}\label{fig:S1}
\end{figure}

Based on the historical data from 1998 to 2020, we implement the univariate functional time series forecasting and functional median polish methods to produce one-step-ahead forecast Lorenz curves at 20 Italian regions. In Figure~\ref{fig:S1}, we display the results for the 20 regions in Italy.

\newpage
\bibliographystyle{plainnat}
\bibliography{Lorenz_curve.bib}

\begin{thebibliography}{52}
\providecommand{\natexlab}[1]{#1}
\providecommand{\url}[1]{\texttt{#1}}
\expandafter\ifx\csname urlstyle\endcsname\relax
  \providecommand{\doi}[1]{doi: #1}\else
  \providecommand{\doi}{doi: \begingroup \urlstyle{rm}\Url}\fi

\bibitem[Ahn and Horenstein(2013)]{AH13}
S.~C. Ahn and A.~R. Horenstein.
\newblock Eigenvalue ratio test for the number of factors.
\newblock \emph{Econometrica}, 81\penalty0 (3):\penalty0 1203--1227, 2013.

\bibitem[Aue et~al.(2015)Aue, Norinho, and H\"{o}rmann]{ANH15}
A.~Aue, D.~D. Norinho, and S.~H\"{o}rmann.
\newblock On the prediction of stationary functional time series.
\newblock \emph{Journal of the American Statistical Association: Theory and
  Methods}, 110\penalty0 (509):\penalty0 378--392, 2015.

\bibitem[Chiou(2012)]{Chiou12}
J.-M. Chiou.
\newblock Dynamical functional prediction and classification with application
  to traffic flow prediction.
\newblock \emph{The Annals of Applied Statistics}, 6\penalty0 (4):\penalty0
  1588--1614, 2012.

\bibitem[Condino(2023)]{Condino23}
F.~Condino.
\newblock Share density-based clustering of income data.
\newblock \emph{Statistical Analysis and Data Mining}, 16:\penalty0 336--347,
  2023.

\bibitem[Dong et~al.(2025)Dong, Shang, Hui, and Bruhn]{DSH+25}
Z.~M. Dong, H.~L. Shang, F.~Hui, and A.~Bruhn.
\newblock A compositional approach to model cause-specific mortality with zero
  counts.
\newblock \emph{Annals of Actuarial Science}, in press, 2025.

\bibitem[Emerson and Hoaglin(1983)]{EH83}
J.~D. Emerson and D.~C. Hoaglin.
\newblock Analysis of two-way tables by medians.
\newblock In D.~C. Hoaglin, F.~Mosteller, and J.~W. Tukey, editors,
  \emph{Understanding robust and exploratory data analysis}. John Wiley \&
  Sons, New York, 1983.

\bibitem[Farris(2010)]{Farris10}
F.~A. Farris.
\newblock {The Gini index and measures of inequality}.
\newblock \emph{The American Mathematical Monthly}, 117\penalty0 (10):\penalty0
  851--864, 2010.

\bibitem[Ferraty and Vieu(2006)]{FV06}
F.~Ferraty and P.~Vieu.
\newblock \emph{{Nonparametric Functional Data Analysis: Theory and Practice}}.
\newblock Springer, New York, 2006.

\bibitem[Gabrys and Kokoszka(2007)]{GK07}
R.~Gabrys and P.~Kokoszka.
\newblock Portmanteau test of independence for functional observations.
\newblock \emph{Journal of the American Statistical Association: Theory and
  Methods}, 102\penalty0 (480):\penalty0 1338--1348, 2007.

\bibitem[Gao et~al.(2019)Gao, Shang, and Yang]{GSY19}
Y.~Gao, H.~L. Shang, and Y.~Yang.
\newblock High-dimensional functional time series forecasting: An application
  to age-specific mortality rates.
\newblock \emph{Journal of Multivariate Analysis}, 170:\penalty0 232--243,
  2019.

\bibitem[Gini(1936)]{Gini36}
C.~Gini.
\newblock On the measure of concentration with special reference to income and
  statistics.
\newblock \emph{Colorado College Publication, General Series}, 208\penalty0
  (1):\penalty0 73--79, 1936.

\bibitem[Gneiting and Katzfuss(2014)]{GK14}
T.~Gneiting and M.~Katzfuss.
\newblock Probabilistic forecasting.
\newblock \emph{Annual Review of Statistics and its Application}, 1:\penalty0
  125--151, 2014.

\bibitem[Gneiting and Raftery(2007)]{GR07}
T.~Gneiting and A.~E. Raftery.
\newblock Strictly proper scoring rules, prediction and estimation.
\newblock \emph{Journal of the American Statistical Association: Review
  Article}, 102\penalty0 (477):\penalty0 359--378, 2007.

\bibitem[Guo et~al.(2022)Guo, Qiao, and Wang]{GQW22}
S.~Guo, X.~Qiao, and Q.~Wang.
\newblock Factor modelling for high-dimensional functional time series.
\newblock Working paper, Renmin University of China, 2022.
\newblock URL \url{https://arxiv.org/abs/2112.13651}.

\bibitem[Hall and Vial(2006)]{HV06}
P.~Hall and C.~Vial.
\newblock Assessing the finite dimensionality of functional data.
\newblock \emph{Journal of the Royal Statistical Society (Series B)},
  68\penalty0 (4):\penalty0 689--705, 2006.

\bibitem[Hao et~al.(2024)Hao, Lin, Wang, and Zhong]{HLW+24}
S.~Hao, S-C. Lin, J-L. Wang, and Q.~Zhong.
\newblock Dynamic modeling for multivariate functional and longitudinal data.
\newblock \emph{Journal of Econometrics}, 239\penalty0 (2):\penalty0 105573,
  2024.

\bibitem[Hooker and Shang(2022)]{HS22}
G.~Hooker and H.~L. Shang.
\newblock Selecting the derivative of a functional covariate in
  scalar-on-function regression.
\newblock \emph{Statistics and Computing}, 32\penalty0 (Article number: 35),
  2022.

\bibitem[Hoover and Yaya(2010)]{HY10}
G.~A. Hoover and M.~E. Yaya.
\newblock {Racial/ethnic differences in income inequality across US regions}.
\newblock \emph{The Review of Black Political Economy}, 37\penalty0
  (2):\penalty0 79--114, 2010.

\bibitem[Horta and Ziegelmann(2018)]{HZ18}
E.~Horta and F.~Ziegelmann.
\newblock {Dynamics of financial returns densities: A functional approach
  applied to the Bovespa intraday index}.
\newblock \emph{International Journal of Forecasting}, 34:\penalty0 75--88,
  2018.

\bibitem[Horv\'{a}th and Kokoszka(2012)]{HK12}
L.~Horv\'{a}th and P.~Kokoszka.
\newblock \emph{{Inference for Functional Data with Applications}}.
\newblock Springer, New York, 2012.

\bibitem[Horv\'{a}th et~al.(2014)Horv\'{a}th, Kokoszka, and Rice]{HKR14}
L.~Horv\'{a}th, P.~Kokoszka, and G.~Rice.
\newblock Testing stationarity of functional time series.
\newblock \emph{Journal of Econometrics}, 179\penalty0 (1):\penalty0 66--82,
  2014.

\bibitem[Hyndman and Khandakar(2008)]{HK08}
R.~J. Hyndman and Y.~Khandakar.
\newblock {Automatic time series forecasting: the forecast package for R}.
\newblock \emph{Journal of Statistical Software}, 27\penalty0 (3):\penalty0
  1--22, 2008.

\bibitem[Hyndman and Shang(2009)]{HS09}
R.~J. Hyndman and H.~L. Shang.
\newblock Forecasting functional time series.
\newblock \emph{Journal of the Korean Statistical Society}, 38\penalty0
  (3):\penalty0 199--211, 2009.

\bibitem[Hyndman et~al.(2024)Hyndman, Athanasopoulos, Bergmeir, Caceres, Chhay,
  O'Hara-Wild, Petropoulos, Razbash, Wang, and Yasmeen]{Hyndman24}
Rob Hyndman, George Athanasopoulos, Christoph Bergmeir, Gabriel Caceres, Leanne
  Chhay, Mitchell O'Hara-Wild, Fotios Petropoulos, Slava Razbash, Earo Wang,
  and Farah Yasmeen.
\newblock \emph{{forecast}: Forecasting functions for time series and linear
  models}, 2024.
\newblock URL \url{https://pkg.robjhyndman.com/forecast/}.
\newblock R package version 8.23.0.

\bibitem[{Jim\'{e}nez-Var\'{o}n} et~al.(2024){Jim\'{e}nez-Var\'{o}n}, Sun, and
  Shang]{JSS23}
C.~F. {Jim\'{e}nez-Var\'{o}n}, Y.~Sun, and H.~L. Shang.
\newblock {Forecasting high-dimensional functional time series: Application to
  sub-national age-specific mortality}.
\newblock \emph{Journal of Computational and Graphical Statistics}, 33\penalty0
  (4):\penalty0 1160--1174, 2024.

\bibitem[Kokoszka et~al.(2019)Kokoszka, Miao, Petersen, and Shang]{KMP+19}
P.~Kokoszka, H.~Miao, A.~Petersen, and H.~L. Shang.
\newblock Forecasting of density functions with an application to
  cross-sectional and intraday returns.
\newblock \emph{International Journal of Forecasting}, 35\penalty0
  (4):\penalty0 1304--1317, 2019.

\bibitem[Kullback and Leibler(1951)]{KL51}
S.~Kullback and R.~A. Leibler.
\newblock On information and sufficiency.
\newblock \emph{The Annals of Mathematical Statistics}, 22\penalty0
  (1):\penalty0 79--86, 1951.

\bibitem[Lai et~al.(2008)Lai, Huang, Risser, and Kapadia]{LHR+08}
D.~Lai, J.~Huang, J.~M. Risser, and A.~S. Kapadia.
\newblock {Statistical properties of generalized Gini coefficient with
  application to health inequality measurement}.
\newblock \emph{Social Indicators Research}, 87:\penalty0 249--258, 2008.

\bibitem[Li et~al.(2020)Li, Robinson, and Shang]{LRS20}
D.~Li, P.~M. Robinson, and H.~L. Shang.
\newblock Long-range dependent curve time series.
\newblock \emph{Journal of the American Statistical Association: Theory and
  Methods}, 115\penalty0 (530):\penalty0 957--971, 2020.

\bibitem[Li et~al.(2024)Li, Li, and Shang]{LLS24}
D.~Li, R.~Li, and H.~L. Shang.
\newblock Detection and estimation of structural breaks in high-dimensional
  functional time series.
\newblock \emph{The Annals of Statistics}, 52\penalty0 (4):\penalty0
  1716--1740, 2024.

\bibitem[{L\'{o}pez-Pintado} and Romo(2009)]{LR09}
S.~{L\'{o}pez-Pintado} and J.~Romo.
\newblock On the concept of depth for functional data.
\newblock \emph{Journal of the American Statistical Association: Theory and
  Methods}, 104\penalty0 (486):\penalty0 718--734, 2009.

\bibitem[Lorenz(1905)]{Lorenz05}
M.~O. Lorenz.
\newblock Methods for measuring the concentration of wealth.
\newblock \emph{Publications of the American Statistical Association},
  9\penalty0 (70):\penalty0 209--219, 1905.

\bibitem[Otto and Salish(2024)]{OS24}
S.~Otto and N.~Salish.
\newblock Approximate factor models for functional time series.
\newblock Working paper, arXiv, 2024.
\newblock URL \url{https://arxiv.org/abs/2201.02532}.

\bibitem[Petersen and M\"{u}ller(2016)]{PM16}
A.~Petersen and H-G. M\"{u}ller.
\newblock {Functional data analysis for density functions by transformation to
  a Hilbert space}.
\newblock \emph{The Annals of Statistics}, 44\penalty0 (1):\penalty0 183--218,
  2016.

\bibitem[{R Core Team}(2024)]{Team24}
{R Core Team}.
\newblock \emph{{R: A Language and Environment for Statistical Computing}}.
\newblock R Foundation for Statistical Computing, Vienna, Austria, 2024.
\newblock URL \url{https://www.R-project.org/}.

\bibitem[Ramsay and Hooker(2017)]{RH17}
J.~O. Ramsay and G.~Hooker.
\newblock \emph{{Dynamic Data Analysis: Modeling Data with Differential
  Equations}}.
\newblock Springer, New York, 2017.

\bibitem[Ramsay and Silverman(2005)]{RS05}
J.~O. Ramsay and B.~W. Silverman.
\newblock \emph{Functional Data Analysis}.
\newblock Springer, New York, 2nd edition, 2005.

\bibitem[Rice and Silverman(1991)]{RS91}
J.~Rice and B.~W. Silverman.
\newblock Estimating the mean and covariance structure nonparametrically when
  the data are curves.
\newblock \emph{Journal of the Royal Statistical Society (Series B)},
  53\penalty0 (1):\penalty0 233--243, 1991.

\bibitem[Shang(2019)]{Shang19}
H.~L. Shang.
\newblock {Visualizing rate of change: An application to age-specific fertility
  rates}.
\newblock \emph{Journal of the Royal Statistical Society: Series A},
  182\penalty0 (1):\penalty0 249--262, 2019.

\bibitem[Shang and Haberman(2020)]{SH20}
H.~L. Shang and S.~Haberman.
\newblock Forecasting age distribution of death counts: {A}n application to
  annuity pricing.
\newblock \emph{Annals of Actuarial Science}, 14:\penalty0 150--169, 2020.

\bibitem[Shang and Haberman(2024)]{SH24}
H.~L. Shang and S.~Haberman.
\newblock Forecasting age distribution of deaths: {C}umulative distribution
  function transformation.
\newblock Working paper, arXiv, 2024.
\newblock URL \url{https://arxiv.org/abs/2409.04981}.

\bibitem[Shang and Haberman(2025)]{SH25}
H.~L. Shang and S.~Haberman.
\newblock Forecasting age distribution of life-table death counts via
  $\alpha$-transformation.
\newblock \emph{Scandinavian Actuarial Journal}, in press, 2025.

\bibitem[Shang et~al.(2022)Shang, Haberman, and Xu]{SHX22}
H.~L. Shang, S.~Haberman, and R.~Xu.
\newblock Multi-population modelling and forecasting life-table death counts.
\newblock \emph{Insurance: Mathematics and Economics}, 106:\penalty0 239--253,
  2022.

\bibitem[Shannon(1948)]{Shannon48}
C.~E. Shannon.
\newblock A mathematical theory of communication.
\newblock \emph{The Bell System Technical Journal}, 27:\penalty0 379--423,
  623--656, 1948.

\bibitem[Sun and Genton(2012)]{SG12}
Y.~Sun and M.~G. Genton.
\newblock Functional median polish.
\newblock \emph{{Journal of Agricultural, Biological and Environmental
  Statistics}}, 17\penalty0 (3):\penalty0 354--376, 2012.

\bibitem[Tang et~al.(2022)Tang, Shang, and Yang]{TSY22}
C.~Tang, H.~L. Shang, and Y.~Yang.
\newblock Clustering and forecasting multiple functional time series.
\newblock \emph{The Annals of Applied Statistics}, 16\penalty0 (4):\penalty0
  2523--2553, 2022.

\bibitem[Tavakoli et~al.(2023)Tavakoli, Nisol, and Hallin]{TNH23}
S.~Tavakoli, G.~Nisol, and M.~Hallin.
\newblock {Factor models for high-dimensional functional time series II:
  Estimation and forecasting}.
\newblock \emph{Journal of Time Series Analysis}, 44\penalty0 (5-6):\penalty0
  601--621, 2023.

\bibitem[Wied(2024)]{Wied24}
D.~Wied.
\newblock Semiparametric distribution regression with instruments and
  monotonicity.
\newblock \emph{Labour Economics}, 90:\penalty0 102565, 2024.

\bibitem[Zhang(2013)]{Zhang13}
J-T. Zhang.
\newblock \emph{Analysis of Variance for Functional Data}.
\newblock Chapman and Hall/CRC, New York, 2013.

\bibitem[Zhang and Wang(2016)]{ZW16}
X.~Zhang and J-L. Wang.
\newblock From sparse to dense functional data and beyond.
\newblock \emph{The Annals of Statistics}, 44\penalty0 (5):\penalty0
  2281--2321, 2016.

\bibitem[Zhou and Dette(2023)]{ZD23}
Z.~Zhou and H.~Dette.
\newblock Statistical inference for high-dimensional panel functional time
  series.
\newblock \emph{Journal of the Royal Statistical Society: Series B},
  85\penalty0 (2):\penalty0 523--549, 2023.

\bibitem[Zizler(2014)]{Zizler14}
P.~Zizler.
\newblock Gini indices and the moments of the share density function.
\newblock \emph{Applications of Mathematics}, 59:\penalty0 167--175, 2014.

\end{thebibliography}

\end{document}